\documentclass{bmcart}

\usepackage[utf8]{inputenc} %unicode support
\usepackage{amsmath,amsfonts,amsthm,bm} % Math packages
\usepackage{tabularx}

\def\includegraphics{}

\startlocaldefs
\endlocaldefs

\usepackage{graphicx}
\usepackage{amsmath}
\usepackage[authoryear]{natbib}
\usepackage[hyperfootnotes=false]{hyperref}
\usepackage{subfig}

\usepackage[final]{changes}   % remove markup
\begin{document}

%%% Start of article front matter
\begin{frontmatter}

\begin{fmbox}
\dochead{Research}

\title{Applications of CryoSat-2 satellite magnetic data in studies of Earth's core field variations}

\author[
addressref={aff1},
email={magdh@space.dtu.dk}
]{\inits{MDH}\fnm{Magnus D.} \snm{Hammer}}
\author[
   addressref={aff1},                   % id's of addresses, e.g. {aff1,aff2}
   corref={aff1},                       % id of corresponding address, if any
%   noteref={n1},                        % id's of article notes, if any
   email={cfinl@space.dtu.dk}   % email address
]{\inits{CC}\fnm{Christopher C.} \snm{Finlay} and \inits{NO}\fnm{Nils} \snm{Olsen}}

\address[id=aff1]{%                           % unique id
  \orgname{Division of Geomagnetism, DTU Space, Technical University of Denmark}, % university, etc
  \street{Centrifugevej 356},                     %
  %\postcode{DK-2800}                                % post or zip code
  \city{Kongens Lyngby},                              % city
  \cny{Denmark\\}                                    % country
}

\begin{artnotes}
%\note{Sample of title note}     % note to the article
%\note[id=n1]{Equal contributor} % note, connected to author
\end{artnotes}

\end{fmbox}% comment this for two column layout

\begin{abstractbox}

\begin{abstract} % abstract
We use 20 years of continuous magnetic field measurements from the {\O}rsted, CHAMP and {\it Swarm} satellite missions, supplemented by calibrated platform magnetometer data from the CryoSat-2 satellite, to study time variations of the Earth's core field at satellite altitude and at the core-mantle boundary (CMB).

From the satellite data we derive composite time series of the core field secular variation (SV) with four-month cadence, at 300 globally distributed Geomagnetic Virtual Observatories (GVO). A previous gap in the GVO series between 2010-2014 is successfully filled using CryoSat-2, and sub-decadal variations are identified. Tests showed that similar sub-decadal SV patterns were obtained from the CryoSat-2 data regardless of whether IGRF-13 or CHAOS-6x9 was used  for the calibration. Cryosat-2 radial field SV series at non-polar latitudes have a mean standard deviation level of 3.5~nT/yr compared to 1.8~nT/yr for CHAMP and 0.9~nT/yr for \textit{Swarm}. GVO radial SV series display regional fluctuations with 5-10 year duration and amplitudes  reaching 20~nT/yr, most notably at low latitudes over Indonesia (2014), over South America and the South Atlantic (2007, 2011 and 2014), and over the central Pacific (2017). 

Applying the Subtractive Optimally Localized Averages (SOLA) method, we map the radial SV at the CMB as a collection of locally-averaged SV estimates.  We demonstrate that using two year Cryosat-2 data windows it is possible to reliably estimate the SV and its time derivative, the secular acceleration (SA), at the CMB, with a spatial resolution of $42^{\circ}$, corresponding to spherical harmonic degree 10. Along the CMB geographic equator, we find strong SA features with amplitude 
$\pm 2.5$\replaced{$\mu\mathrm{T}/\mathrm{yr}^2$}{$\mathrm{nT}^2/\mathrm{yr}^2$} under Indonesia from 2011-2014, under central America from 2015-2019, and sequences of alternating SA under the Atlantic during 2004-2019.  We find that platform magnetometer data from CryoSat-2 makes a valuable contribution to the emerging picture of sub-decadal core field variations.

Using one year {\it Swarm} data windows, we show that it is possible to study SA changes at low latitudes on timescales down to 1 year, with spatial resolution of $42^{\circ}$. We find strong positive and negative SA features appearing side-by-side in the Pacific in 2017, and thereafter drift westward.

\end{abstract}
%%%%%%%%%%%%%%%%%%%%%%%%%%%%%%%%%%%%%%%%%%%%%%
%%                                          %%
%% The keywords begin here                  %%
%%                                          %%
%% Put each keyword in separate \kwd{}.     %%
%%                                          %%
%%%%%%%%%%%%%%%%%%%%%%%%%%%%%%%%%%%%%%%%%%%%%%

\end{abstractbox}

\end{frontmatter}

%%%%%%%%%%%%%%%%%%%%%%%%% start of article main body
\section{Introduction}
\label{sec:Intro}
The main part of the Earth's magnetic field is generated by motions in the electrical conducting liquid outer core, in a process known as the geodynamo. This magnetic field, termed the core field, exhibits both spatial and temporal changes over a broad range of scales. Magnetic measurements from satellites have increased the recovery of small-scale features of this field and revealed rapid changes in its temporal behaviour \replaced{\cite[e.g.][]{Alken_etal_2020b,Finlay_etal_2020,Baerenzung_etal_2020,Ropp_etal_2020,Sabaka_etal_2020}} {\mbox{\citep{Olsen_Mandea_2008,Alken_etal_2020b,Finlay_etal_2020}}}. From ground and space magnetic observations, variations in the first and second time derivatives of the field, termed the secular variation (SV) and acceleration (SA), respectively, may now be resolved down to \replaced{periods of about 1 to 2 years} { a period of about 1 year} \replaced{\mbox{\citep{Lesur_etal_2010,Lesur_etal_2017,Ropp_etal_2020}}}{\mbox{\citep{Chulliat_Maus_2014}}}.

% SV, SA, physical mechanisms behind 
Studies have revealed oscillating SA pulse-like field features at the core-mantle boundary (CMB) focused in the region around the geographical equator \citep{Chulliat_etal_2010,Chulliat_Maus_2014,Chulliat_etal_2015,Sabaka_etal_2018,Alken_etal_2020b}. The interpretation and geophysical mechanisms responsible for driving such distinctive behaviour in the SA signal remains under debate \citep[e.g.][]{Gillet_2019,Buffett_Matsui_2019, Aubert_Finlay_2019, Gerick_etal:2020}, as is the connection to abrupt changes in the SV observed at ground observatories \citep{Mandea_etal_2010}. \replaced{The secular acceleration must be characterized with care,}{Characterizing the secular acceleration should be done with care,} \added{paying attention to those spatial and temporal scales that are well resolved,} as its \added{observed} spatial spectra at the core surface is blue, showing increasing power with spherical harmonic degree, and \replaced{its observed temporal spectra seems to be rather flat, meaning that there could be important unresolved fast variations}{ an instantaneous interpretation may be formally undefined as the time window used to estimate it goes towards zero} \citep{Bouligand_etal_2016,Lesur_etal_2017,Gillet_2019,Christensen_etal_2012}. In this \replaced{respect, assessing}{ aspect, accessing} the limitations of the \added{information that may be obtained from the} measurements by analysing their resolving power becomes \replaced{crucial}{ important}  when \replaced{seeking to investigate}{ trying to recover} the SA signal at small length scales and short time scales.

Magnetic field measurements from low Earth orbiting (LEO) satellites provide global observations of the field, which have proved important for mapping the spatial structures of the core field signal \citep{Olsen_Stolle_2012}. Since the launch of the Danish {\O}rsted (1999-2014) satellite, the German CHAMP (2000-2010) satellite and the European {\it Swarm} (2013-) satellite trio, satellites have provided high quality magnetic field measurements, which have enable investigations into the spatio-temporal variations of field. Unfortunately, the CHAMP mission ended in September 2010 and since reliable vector measurements from the {\O}rsted satellite extend only up to 2006, there is a gap from 2010 and 2014 in the satellite magnetic records \citep{Finlay_etal_2016a}. However, other satellite missions, not dedicated to measuring the magnetic field, may offer a possibility to fill in this gap adding information about the field. In particular, the CryoSat-2 (2010-) mission, intended for measuring polar ice thickness, carries three platform magnetometers for navigational purposes. Calibrated CryoSat-2 measurements, where vector fluxgate magnetometer readings have been transformed into reliable magnetic field vector outputs, from August 2010 to December 2018 are now available \citep{Olsen_etal_2020}, such that there are in total 20 years of continuous satellite measurements. With the data from CryoSat-2 now available, it is important to assess the quality and limitations of such calibrated platform magnetometer measurements and to test what contribution they can make to studying core field variations. 

The standard approach for using satellite magnetometer data for core field studies is to construct spherical harmonic (SH) field models by  least-squares inversion methods \cite[see e.g.][]{Langel_1987}. In such global models B-splines are often used to parameterize the model time-dependence\added{.} Typically this necessitates temporal regularization which modifies the time-dependence of the harmonics in a non-uniform manner. An undesirable consequence is that for the higher harmonics, the first time derivative effectively becomes a time average over an increasingly long interval rather than an estimate of the instantaneous secular variation \citep{Olsen_etal_2009}. \added{Cryosat-2 data has already been used in the construction of such time-dependent spherical harmonic field models by \mbox{\citet{Alken_etal_2020b, Finlay_etal_2020, Kloss_etal_2021}}.} In an effort to access more detailed information on the spatial and temporal structure of the field, it is also of interest to look at alternative techniques for studying secular variation that can complement the traditional SH approach. In this paper we focus on two different local methods for studying core field variations as recorded in satellite measurements, with a particular focus on assessing the quality and resolving ability of CryoSat-2 magnetic data.

In a first assessment of the quality of the CryoSat-2 data and its ability to map the SV field at satellite altitude together with CHAMP and {\it Swarm} we use the \replaced{Geomagnetic}{Geomagentic} Virtual Observatory (GVO) technique developed by \cite{Mandea_Olsen_2006,Olsen_Mandea_2007}. This technique involves computing time series of field estimates at specified target locations at satellite altitude, from satellite measurements taken nearby. We apply the processing algorithm recently developed to derive \added{four-monthly} \textit{Swarm} GVO data series \citep{Hammer_etal_2020} and derive time series on a global grid of 300 GVOs. This \deleted{global} network of GVOs allows field changes at satellite altitude to be investigated globally \added{at fixed locations. The GVOs provide a useful compression of satellite magnetic measurements and are a convenient dataset for workers wishing to use constraints for studies of core dynamics. They provide an alternative to fitting core dynamics models to conventional spherical harmonic field models. They involve a series of independent local constraints rather than global constraints, and have well understood error covariances that can be assigned by methods similar to those used with ground observatories. GVOs have already been used by a number of groups for studies of core dynamics \mbox{\cite[e.g.][]{Whaler_Beggan_2015,Barrois_etal_2018,Domingos_etal_2019}}. 
%In studies of core dynamics, fitting to independent local GVOs is an alternative to fitting to spherical harmonic coefficients representing global functions.  GVOs have relatively simple error budgets that are assigned by methods similar to those used for ground observatories.    They have previously been used by a number of groups for studies of core processes \cite[e.g.][]{Whaler_Beggan_2015, Domingos_etal_2019, Barrois_etal_2018}.
}

\replaced{In the second part of this study we directly map}{Going further, we seek to map} the radial field SV  at the core-mantle boundary, using the technique of Subtractive Optimally Localized Averages (SOLA) that was adapted to geomagnetism by \cite{Hammer_Finlay_2019}. The SOLA technique can be used to compute estimates of the radial field SV directly at the CMB, based on local spatial averages described by averaging kernels, of the SV field centred on target locations of interest, and time-averaged over chosen time windows. By collecting many individual SOLA estimates on a grid at the CMB, the SV field can be mapped on regional or global scales. \replaced{An important foundation of the SOLA technique is that for noise-free data and a linear forward problem, any linear combination of the data provides a specific average of the true model. With noisy data, a variance is ascribed to this spatial average value such that a trade-off between resolution and variance arises \mbox{\citep{Oldenburg_1984,Parker_1994}}.}{An important aspect of the SOLA technique, is that the estimated averaged field is the only unique information provided by the satellite measurements; that is, the estimate along with the averaging kernel constitutes our knowledge of the field in the vicinity of the target location \mbox{\citep{Oldenburg_1984,Parker_1994}}.} The SOLA technique readily provides information on the resolution offered by the magnetic field observations, in the form of averaging kernels, \replaced{and}{as well as} estimates of the variance of the locally averaged field. We compare SOLA-based maps of SV and SA estimates derived from CryoSat-2 and {\it Swarm} data, in order to asses the possibilities of the CryoSat-2 data for mapping these fields at the CMB. Demonstrating the usefulness of the CryoSat-2 data, we then take advantage of this data to map the time evolution of the SA field along the geographic equator at the CMB from 2001 to 2019. 

\added{We wish to emphasize that both the GVO and the SOLA methods can result in patterns of secular variation different from those seen in the CHAOS field model, despite the fact they use similar data selection schemes and the same magnetospheric field model. In the GVO and SOLA methods data close to a location of interest are used to effectively determine localized field or SV estimates.  In contrast, estimation of the coefficients of truncated spherical harmonic expansions, in models such as CHAOS, is an inherently global approach that aims to find the best possible global model. In the GVO method only data from within a 700km radius of a target location is used to determine the local potential.  In the SOLA method measurements far from the site of interest have essentially no influence on the estimated CMB SV because the data kernels, based on Green's functions for Laplace's equation under Neumann boundary conditions, have decreasing sensitivity far away from the target location. 

Moreover, the CHAOS model involves temporal regularization whereby one minimizes global norms based on time derivatives of the CMB radial field, integrated over the entire timespan of the model. In the GVO method there is no temporal smoothing beyond the choice of 4 month data windows and use of annual differences to produce SV series. In the SOLA method we use one or two year windows to estimate the SV and then annual differences to estimate SA. The SOLA method involves a trade-off parameter specifying the balance between the spatial resolution and the variance of each local estimate, each estimate being time-averaged over a one or two year time window; there is no global regularization over longer time spans.  Both methods therefore constitute a localized compression of information contained within the satellite data on the potential field near the location of interest within the specified time window.  The GVOs and SOLA can thus give a different picture of the core field evolution compared to the spherical-harmonic based CHAOS model with its global temporal regularization, especially regarding rapid changes at short wavelengths scales when the data quality is high (see Section \ref{sec:SOLA_results}).}

Section \ref{sec:data} describes the satellite measurements used in this study, including the CryoSat-2 platform magnetometer data, and how these have been selected and processed. Section \ref{sec:GVO} presents the GVO technique and results concerning global GVO time series.  Section \ref{sec:SOLA} presents the SOLA technique and results of applying this to estimate SV and SA at the CMB. Conclusions and perspectives are given in Section \ref{sec:Conclusions}.

\section{Data Selection}
\label{sec:data}
We use satellite vector magnetic field measurements from the {\O}rsted satellite between July 2000 and December 2005, from CHAMP taking L3 magnetic data between July 2000 and September 2010, and from the {\it Swarm} trio taking Level 1b MAG-L data, version 0505/0506, between January 2014 and April 2020. Most importantly for this study we make use of platform magnetometer data from the CryoSat-2 mission, taking calibrated vector measurements with a sampling rate of 4s from the FGM1 magnetometer dataset, version 3, between August 2010 to December 2018.  This dataset has had extensive corrections applied for disturbances fields, and was calibrated using a reference field model - for full details see \citet{Olsen_etal_2020}. 

From the {\O}rsted, CHAMP and {\it Swarm} measurements we produced two data sets: dataset \#1 used in the GVO application taking a 15~s subsampling of the vector field measurements from {\O}rsted, CHAMP and {\it Swarm}, while taking every 4th measurement from the CryoSat-2 dataset (i.e. a 16~s sub-sampling); dataset \#2 used in the SOLA application taking a 5~s subsampling of the vector field measurements from {\O}rsted, CHAMP and {\it Swarm}, while taking every element of the CryoSat-2 dataset with its 4~s sampling rate. Field measurements having gross data outliers for which the vector field components deviated more than 500~nT from the CHAOS-7.2 internal field model predictions \citep{Finlay_etal_2020} were rejected. For both data sets we apply a dark geomagnetically quiet time selection criteria scheme.  See table~\ref{table:1} below for full details of the selection requirements for the datasets used in the GVO and SOLA applications. In both cases we required the sun to be at least $10^{\circ}$ below the horizon, adding restrictions on the geomagnetic activity index $K_p$ and the change in the ring current index \citep[see][]{Olsen_etal_2014}, as well as constraints on the merging electric field at magnetopause, and on the magnitudes of the $B_Y$ and $B_Z$ components of the interplanetary magnetic field \citep[e.g.][]{Finlay_etal_2020, Ritter_etal_2004}. We used minute values of the IMF components and solar wind speed from the OMNI database, \url{http://omniweb.gsfc.nasa.gov}, computing two hourly means prior to the time of the considered datum \citep{Finlay_etal_2016a}. 

\added{Since our focus is the core field we have applied corrections to the data for the lithospheric and external fields.  For the lithospheric field model, we used the LCS-1 model \citep{Olsen_etal_2017}; the precise choice of lithospheric field is not crucial for studies of the SV and SA. For the solar-quiet ionospheric field and associated induced fields we used the CIY4 model \citep{Sabaka_etal_2018}.  For the magnetospheric field and related induced fields we used the CHAOS-7 model  \citep{Finlay_etal_2020}.  These models were chosen as they are well established and compatible with the data selection criteria described above.}  

\begin{table}[!h]
\centering
\centerline{\begin{tabularx}{1.0\linewidth}{l | l X l}
\hline
                             & \textit{Data set \# 1}      && \textit{Data set \# 2} \\
\hline
\textit{Used in}             & \textit{GVO application}  && \textit{SOLA application} \\
\textit{Satellite}           & CHAMP, CryoSat-2, {\it Swarm}  && {\O}rsted, CHAMP, CryoSat-2, {\it Swarm}  \\
\textit{Data type}           & \textit{vector data sums and differences}     && \textit{radial vector data}    \\
\textit{Subsampling}         & 15~s  (16~s CryoSat-2)                     &&  5~s (4~s CryoSat-2)           \\
\textit{Kp}                  & $<3^o$                   &&  $<2^o$ \\
$\vert dRC/dt\vert$          & $<3$~nT/h                 &&  $<2$~nT/h  \\
\textit{Em}                  & $\leq0.8$~mV/m            &&  $\leq0.8$~mV/m \\
IMF $B_Z$                    & $>0$~nT                   &&  $>0$~nT \\
IMF $\vert B_Y \vert$        & $<10$~nT                  &&  $<6$~nT \\
\textit{Solar angle}         & $<-10^{\circ}$           &&  $<-10^{\circ}$ \\
\textit{Outliers removed}    & $<500$~nT from CHAOS      &&  $<500$~nT from CHAOS \\
\hline
\textit{LCS-1}               &  \multicolumn{3}{c}{Lithospheric field for $n \in [14,120]$ subtracted} \\
\textit{CHAOS-7.2}           &  \multicolumn{3}{c}{Magnetospheric (plus induced) fields subtracted} \\
\textit{CIY4}                &  \multicolumn{3}{c}{Ionospheric (plus induced) fields subtracted} \\
\hline
\end{tabularx}}
\caption{Selection criteria, and model corrections applied for the GVO and SOLA datasets used in this study.}
\label{table:1}
\end{table} 

As noted in Table \ref{table:1}, for the GVO application we use the sums and differences of the magnetic field measurements. \added{It has been shown by \mbox{\citet{Olsen_etal_2015}} and \mbox{\citet{Sabaka_etal_2018}}, that taking differences of the satellite measurements along-track and east-west (between \textit{Swarm} satellites A and C) helps the recovery of the small scale core field, as this reduces the impact of correlated errors caused by unmodelled large-scale external fields. Here we follow such an approach by taking differences of the measurements, but we also include along-track and east-west sums of the measurements in order to ensure sufficient constraint on the larger wavelengths of the field \mbox{\citep{Sabaka_etal_2013,Hammer_2018}}.} We denote the magnetic vector measurements by $B_k(\mathbf{r})$, where $k$ is any \replaced{o}{p}f the three given vector component of the field, $\Delta d_k$ and $\Sigma d_k$ denote measurement differences and sums of this particular component, respectively. Here the along-track (AT) and East-West (EW) data differences are denoted by $\Delta d_k=(\Delta d_k^{\mathrm{AT}},\Delta d_k^{\mathrm{EW}})$, and the data sums by $\Sigma d_k=(\Sigma d_k^{\mathrm{AT}},\Sigma d_k^{\mathrm{EW}})$. The along-track data differences are calculated using the 15~s differences $\Delta d_k^{\mathrm{AT}} = [B_k(\mathbf{r},t) - B_k(\mathbf{r}+\delta \mathbf{r},t+15s)]$. The along-track sums were calculated as $\Sigma d_k^{AT} = [B_k(\mathbf{r},t) + B_k(\mathbf{r}+\delta \mathbf{r},t+15s)]/2$. For \textit{Swarm}, East-West differences were calculated as $\Delta d_k^{\mathrm{EW}} = [B_k^{\mathrm{SWA}}(\mathbf{r}_1,t_1) - B_k^{\mathrm{SWC}}(\mathbf{r}_2,t_2)]$ having an East-West orbit separation between the \textit{Swarm} \textit{Alpha} (SWA) and \textit{Charlie} (SWC) satellites of $\approx 1.4^{\circ}$ corresponding to 155~km at the equator \citep{Olsen_etal_2015}. The East-West sums were calculated as $\Sigma d_k^{\mathrm{EW}} = [B_k^{\mathrm{SWA}}(\mathbf{r}_1,t_1) + B_k^{\mathrm{SWC}}(\mathbf{r}_2,t_2)]/2$. For a particular orbit of \textit{Swarm} \textit{Alpha} the corresponding \textit{Swarm} \textit{Charlie} data were selected to be those closest in colatitude with the condition that $\vert\Delta t\vert=\vert t_1-t_2\vert<50s$ \citep{Olsen_etal_2015}. 

\section{Application I: Geomagnetic Virtual Observatories}
\label{sec:GVO}

\subsection{Four monthly Core Field GVOs and Secular Variation Estimates}
\label{sec:VO_method}
In the first application, we compute Geomagnetic Virtual Observatory time series derived from dataset \#1. Of particular interest is the quality of the GVO series obtained from CryoSat-2 data compared with similar series obtained from CHAMP and \textit{Swarm} data. The time series consist of estimates of the geocentric spherical polar vector components of the magnetic field at specified target points, referred to as GVOs \citep{Mandea_Olsen_2006,Olsen_Mandea_2007}. Here we use the same algorithm described in detail by \cite{Hammer_etal_2020} (see also \url{http://www.spacecenter.dk/files/magnetic-models/GVO/GVO_Product_Algorithm.pdf}) to produce the \textit{Swarm} GVO product, and derive global grids of 300 uniformly distributed GVO time series each having 4 month cadence. The GVOs are located in an approximately equal area grid based on the sphere computed using the algorithm of \cite{Leopardi_2006}. 

For each GVO in the grid we take data from within a cylinder of horizontal radius $r_{cyl}=700$\,km. The GVO's have the spherical polar coordinates $\mathbf{r}_{GVO}=(r,\theta,\phi)$, and are placed at fixed altitudes $r=r_a+h_{GVO}$ where $h_{GVO}$ is the height above the Earth's mean spherical radius $r_a=6371.2$\,km. For the CHAMP, CryoSat-2 and \textit{Swarm} missions the GVO altitudes were chosen as $h_{GVO}=370$~km, $h_{GVO}=727$~km and $h_{GVO}=490$~km, respectively, such that the GVO's are located at approximately the mean orbital altitude for each mission during the time interval considered. 

The input measurements of dataset \#1 are provided in an Earth-Centred-Earth-Fixed (ECEF) coordinate frame by the spherical polar components $\mathbf{B}^{obs}=(B_r,B_{\theta},B_{\phi})$. From the extracted vector field measurements surrounding each GVO, within a radius of 700\,km and within a 4 month time window, a residual magnetic field, $\delta \mathbf{B}$, is first computed by subtracting off estimates of the main field and non-core fields 
\begin{equation}
\delta \mathbf{B} = \mathbf{B}^{obs} - \mathbf{B}^{MF} -\mathbf{B}^{lith}- \mathbf{B}^{mag}- \mathbf{B}^{iono}   \label{eq:res}
\end{equation}
where the field estimates removed are: a) $\mathbf{B}^{MF}$ the internal field for SH degrees $n \in [1,13]$ as given by IGRF-13 \citep{Alken_etal_2020}, b) $\mathbf{B}^{lith}$ the static internal field for SH degrees $n \in [14,185]$ as given by the LCS-1 model, \citep{Olsen_etal_2017}, c) $\mathbf{B}^{mag}$ the magnetospheric and associated induced field as given by the CHAOS-7.2, model, \citep{Finlay_etal_2020}, d) $\mathbf{B}^{iono}$ the ionospheric and associated induced field as given by the CIY4 model, \citep{Sabaka_etal_2018}. Note that estimates of the main field from IGRF-13 (with linear time dependence over 5 year intervals) are subtracted here. At a later stage, main field estimates, again from IGRF-13, are added back for the specified GVO times and positions. Removal of a main field at this stage allows for a more effective pre-whitening of the data, such that Huber weights, used in the robust GVO estimation scheme, can be well determined.  We emphasize that this approach still allows us to capture departures from the subtracted main field when required by the satellite data. 

\added{Of the remaining residual field, we are interested in the core field part of that signal. Although we have removed estimates of the external fields and their associated Earth-induced counterpart, contributions from non-core sources remain in the residual field. A particular concern is contamination from fields caused by rapidly varying ionospheric currents, for example polar electrojet currents at high latitudes in the E-layer, and possibly also signatures of F-layer currents at mid and low latitudes. Further work is needed on these aspects.  We attempt to mitigate leakage of field-aligned currents by removing toroidal field estimates \citep{Sabaka_etal_2010} obtained by performing an epoch-by-epoch spherical harmonic analysis performed on the global network of GVOs \citep{Hammer_etal_2020}, see below for further details.}

Next, the residual magnetic field data and their positions are transformed from the spherical system to a right-handed local topocentric Cartesian system $(x,y,z)$ with origin at the GVO target location. At the GVO location and only at this location, $x$ points towards geographic south, $y$ points towards east and $z$ points radially upwards \citep{Hammer_etal_2020}. 
Assuming that the satellite measurements are made in a source free region, the residual magnetic field, $\delta \mathbf{B}$, is a Laplacian potential field. In the local Cartesian coordinate system the magnetic scalar potential, $V$, can be expanded as a sum of polynomials of the form $C_{abc} x^a y^b z^c$    \citep{Backus_etal_1996}.  Here we use this expansion out to cubic terms

\begin{align}
V(x,y,z) &
= C_{100}x + C_{010}y + C_{001}z + C_{200}x^2 + C_{020}y^2 + C_{002}z^2            
 \nonumber \\
         &
+ C_{110}xy + C_{101}xz + C_{011}yz + C_{300}x^3 + C_{030}y^3 + C_{003}z^3
 \nonumber \\
         &
+ C_{210}x^2y+ C_{201}x^2z  + C_{120}y^2x  + C_{021}y^2z + C_{102}z^2x + C_{012}z^2y  \nonumber \\
 &
 + C_{111}xyz 
         \label{eq:pot_backus}
\end{align}

The forward problem linking the vector of GVO model coefficients, $\mathbf{m}=[C_{100},C_{010},..., C_{111}]^T$, with the data vector $\mathbf{d}^{vec}$  containing the residual field components, $\delta \mathbf{B}$, can be written

\begin{equation}
\mathbf{d}^{vec} = \underline{\underline{\mathbf{G}}}^{vec} \mathbf{m} \label{eq:GVO_forward}
\end{equation}

where $\underline{\underline{\mathbf{G}}}^{vec}$ is a design matrix derived from the spatial derivatives of eq.~\eqref{eq:pot_backus}. 
As noted above, instead of using residual vector field components to compute the potential, we use sums and differences of the residual field vector components such that the data vector is $\mathbf{d}=\{\Delta d_{x}^{vec},\Delta d_{y}^{vec},\Delta d_{z}^{vec},\Sigma d_{x}^{vec},\Sigma d_{y}^{vec},\Sigma d_{z}^{vec}\}$, where $\Delta$ and $\Sigma$ denotes the differences and sums of the computed residual field as described in Section \ref{sec:data}. The relevant design matrix is then constructed as $\underline{\underline{\mathbf{G}}}=\{\Delta G_x^{vec},\Delta G_y^{vec},\Delta G_z^{vec},\Sigma G_x^{vec},\Sigma G_y^{vec},\Sigma G_z^{vec})\}$ where $\Delta G_{k}^{vec} = [G_{k}^{vec}(\mathbf{r}_1) - G_{k}^{vec}(\mathbf{r}_2)]$ and $\Sigma G_{k}^{vec} = [G_{k}^{vec}(\mathbf{r}_1) + G_{k}^{vec}(\mathbf{r}_2)]/2$ where $k=(x,y,z)$. 

To determine the GVO model coefficients, we use a robust iteratively reweighted least-squares inversion scheme, based on a diagonal weight matrix consisting of robust (Huber) weights for each entry in the data vector \cite[e.g.][]{Constable_1988}.  We also \replaced{include}{including} a down-weighting factor of 1/2 for the \textit{Swarm} satellites Alpha and Charlie accounting for the fact that these two satellites fly side-by-side and therefore do not provide completely independent measurements. A minimum number of 30 data points were required for the computing the inversion. Using the resulting coefficients of the potential for a given GVO target location and time, derived from the associated sums and differences satellite data, a prediction for the mean residual field at the GVO target point and epoch can be computed as $\delta \mathbf{B}_{GVO}(x,y,z)=-\nabla V(0,0,0)=-(C_{100},C_{010},C_{001})$. 

Moving back to the vector components in spherical polar \replaced{coordinates}{coordiantes}, $\delta B_{GVO,r}=\delta B_{GVO,z}$, $\delta B_{GVO,\theta}=\delta B_{GVO,x}$, $\delta B_{GVO,\phi}=\delta B_{GVO,y}$.  We then add back the IGRF-13 main field predictions for the given target point and epoch, $\mathbf{B}_{GVO}^{MF}(\mathbf{r}_{GVO},t)$ to obtain

\begin{equation}
\mathbf{B}_{GVO}(\mathbf{r}_{GVO},t) = \delta\mathbf{B}_{GVO}(\mathbf{r}_{GVO},t) + \mathbf{B}_{GVO}^{MF}(\mathbf{r}_{GVO},t) \label{eq:5}
\end{equation}

The above procedure is then repeated for each epoch and each component to obtain time series of GVO estimates of the  vector magnetic field at the GVO target locations. 

\replaced{The GVO method assumes that the residual field in eq.\eqref{eq:res}, is a potential field. However, because the satellite measurements are made in the ionospheric F-region, in-situ currents can cause non-potential fields to leak into the estimated potential \mbox{\citep{Olsen_Mandea_2007}}. Therefore, in}{In} a final post-processing step, we carry out an epoch-by epoch spherical harmonic analysis of our 4-monthly GVOs, estimating external and toroidal field contributions to SH degree 13 or to degree 6 if few\added{er} than 300 GVOs are available. For epochs having an insufficient number of GVOs available to ensure a stable solution, the external and toroidal coefficients were computed by a linear interpolation between nearby epochs. The external and toroidal field estimates\added{, reaching a level of $\pm15$nT at high latitudes,} are then removed to obtain an estimated Core Field GVO time series \citep{Hammer_etal_2020}. 

Secular variation at a given GVO location is computed using annual differences between values at time $t+6$~months and at time $t-6$~months. \added{We have chosen to take annual differences, as this helps to avoid annual non-core signals that may persists in the GVO series despite our best efforts in reducing such contamination.}

\begin{figure*}[!h]
\centerline{\includegraphics[angle=0, width=\textwidth]{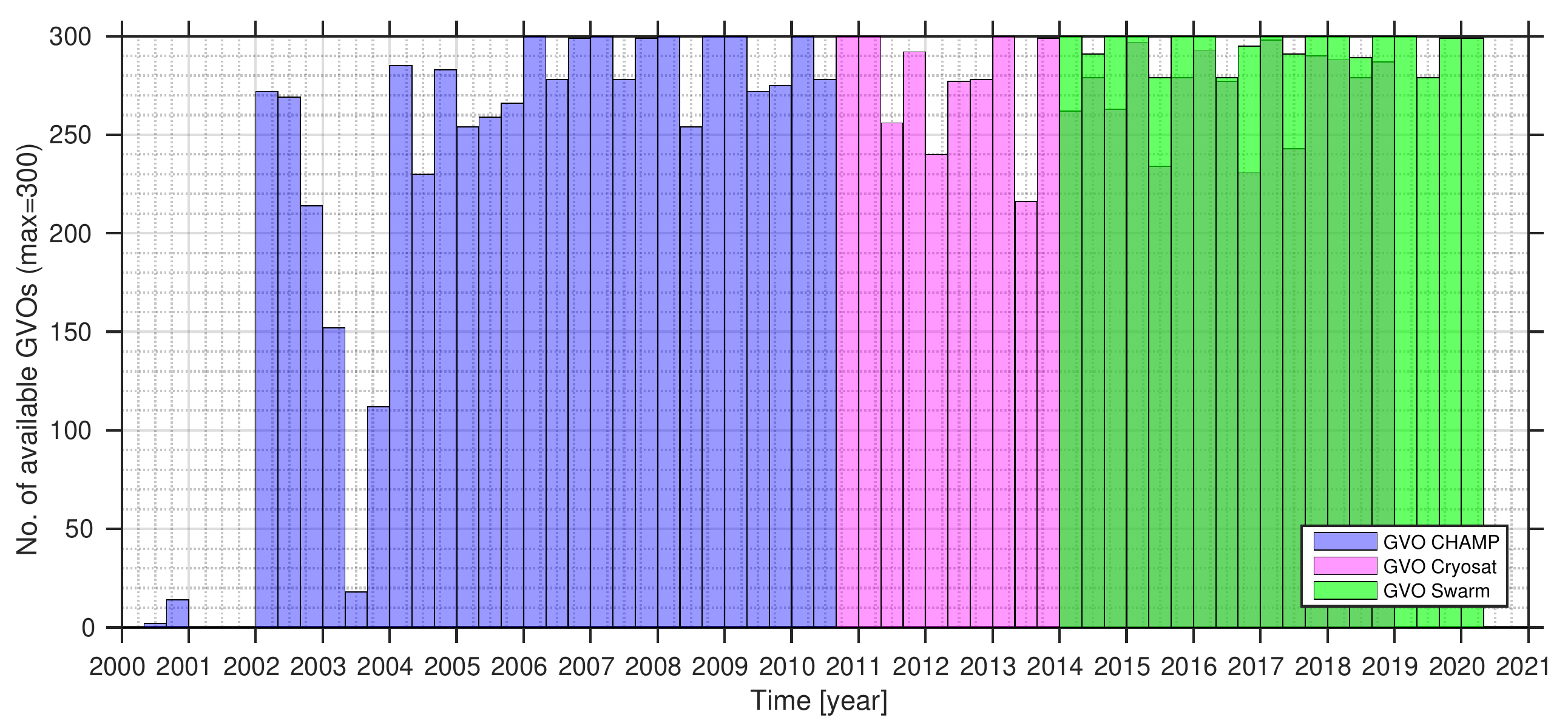}}
\caption{The available number of GVO's for each epoch during CHAMP (purple), CryoSat-2 (pink) and {\it Swarm} (green).}
\label{Fig:2}
\end{figure*}

\added{Figure~\ref{Fig:2} presents the number of 4-monthly Core Field GVO estimates during the past 20 years.  The maximum possible number of GVOs per epoch is 300. A strong dip in the number of GVOs is seen during 2002-2004 due to} \deleted{the intermittent availability of CHAMP vector data during this time as well as} \added{increased solar activity that meant there were fewer data meeting our selection criteria. As noted above, if fewer than 30 measurements are available within a GVO target cylinder during a given 4-month window, we were unable to \replaced{reliably}{reliable} determine GVO estimates.  The remaining epochs from 2004-2020 are well covered with only few epochs having less than 250 GVO's available. We find that using CryoSat-2 data, we  are able to provide between 200 and 300 GVO estimates at all times during the gap between the end of the CHAMP mission and the start of the \textit{Swarm} mission.}

%******************************************
%******************************************

\subsection{GVO Results: Global time series of Secular Variation from 2002 to 2020}
\label{sec:VO_results}
\deleted{Figure~\ref{Fig:2} presents the number of 4-monthly Core Field GVO estimates during the past 20 years.  The maximum possible number of GVOs per epoch is 300. A strong dip in the number of GVOs is seen during 2002-2004 due to the intermittent availability of CHAMP vector data during this time as well as increased solar activity that meant there were fewer data meeting the selection criteria. As noted above, if fewer than 30 measurements are available within a GVO target cylinder during a given 4-month window, we were unable to reliable determine GVO estimates.  The remaining epochs from 2004-2020 are well covered with only few epochs having less than 250 GVO's available.  We find that using CryoSat-2 data, we  are able to provide between 200 and 300 GVO estimates at all times during the gap between the end of the CHAMP mission and the start of the \textit{Swarm} mission.} 

Table \ref{table:2} presents the root-mean-square (rms) and mean of the residuals between the contributing satellite data (sums and differences) and GVO model predictions, summing over all GVO's for a given vector component and a given region (polar or nonpolar). Here we defined polar to be polweard of $\pm 54^{\circ}$ geographic latitude. The polar rms values for both sums and differences are higher than the non-polar, and the CHAMP values are slightly higher than the \textit{Swarm} values. The CryoSat-2 values are seen to be higher for all components but not unreasonable, given they are derived from platform magnetometer data. The non-polar rms values for all components are below 2~nT during both CHAMP and \textit{Swarm} times. The CryoSat-2 GVO's rms values are as expected larger, and especially for the along-track differences, indicating that along-track correlated noise is less dominant, due to the presence of other noise sources in platform magnetometer data.  	

%******************************************
\begin{table}[!h]
\centering
{\scriptsize
\begin{tabular}{l l  |  l  r  r | l  r  r| l  r  r}
\hline 
&           & \multicolumn{3}{c|}{CHAMP} & \multicolumn{3}{c|}{CryoSat-2} &  \multicolumn{3}{c}{{\it Swarm}} \\
%& Component & No. & Mean [nT] & rms [nT]  & No. & Mean [nT] & rms [nT] & No. & Mean [nT] & rms [nT] \\
& Component & No. & Mean  & rms   & No. & Mean  & rms & No. & Mean  & rms  \\
&           &     &   [nT] &   [nT]  &   &   [nT] &   [nT] &  &  [nT] &  [nT] \\
\hline 
Polar           &                 & 2574   &       &      & 2106 &      &      & 1872 &        &   \\

                & $\sum B_{x,NS}$ &        & -0.01 & 6.61 &      & 0.01 & 8.21 &      & 0.02   & 6.39  \\
                & $\sum B_{y,NS}$ &        & 0.00  & 6.52 &      & 0.00 & 8.37 &      & -0.02  & 7.05  \\
                & $\sum B_{z,NS}$ &        & 0.00  & 3.34 &      & 0.00 & 4.11 &      & -0.03  & 3.12  \\
                & $\sum B_{x,EW}$ &        &       &      &      &      &      &      & 0.04   & 6.06  \\
                & $\sum B_{y,EW}$ &        &       &      &      &      &      &      & 0.01   & 6.67  \\
                & $\sum B_{z,EW}$ &        &       &      &      &      &      &      & 0.05   & 2.94  \\                                             
                & $\Delta B_{x,NS}$ &      & 0.00  & 4.35 &      & 0.01 & 7.89 &      & 0.01   & 3.84  \\
                & $\Delta B_{y,NS}$ &      & -0.01 & 5.20 &      & 0.01 & 8.01 &      & 0.00   & 4.94  \\
                & $\Delta B_{z,NS}$ &      & 0.01  & 1.61 &      & -0.00& 4.85 &      & 0.00   & 1.39  \\
                & $\Delta B_{x,EW}$ &      &       &      &      &      &      &      & 0.16   & 3.22  \\
                & $\Delta B_{y,EW}$ &      &       &      &      &      &      &      & 0.08   & 3.43  \\
                & $\Delta B_{z,EW}$ &      &       &      &      &      &      &      & -0.05  & 0.96  \\                        
\hline   
Non-polar       &                 & 7326   &       &      & 5994 &      &      & 5328 &        &   \\
                & $\sum B_{x,NS}$ &        & -0.01 & 1.76 &      & -0.01& 3.85 &      & 0.00   & 1.76  \\
                & $\sum B_{y,NS}$ &        & 0.00  & 1.47 &      & 0.00 & 3.31 &      & -0.01  & 1.95  \\
                & $\sum B_{z,NS}$ &        & 0.00  & 1.33 &      & 0.00 & 3.08 &      & -0.00  & 1.08  \\
                & $\sum B_{x,EW}$ &        &       &      &      &      &      &      & -0.04  & 1.65  \\
                & $\sum B_{y,EW}$ &        &       &      &      &      &      &      & 0.02   & 1.66  \\
                & $\sum B_{z,EW}$ &        &       &      &      &      &      &      & -0.02  & 0.99  \\                                             
                & $\Delta B_{x,NS}$ &      & -0.01 & 0.50 &      & -0.01& 4.78 &      & 0.00   & 0.27  \\
                & $\Delta B_{y,NS}$ &      & 0.00  & 0.58 &      & 0.01 & 4.36 &      & 0.00   & 0.38  \\
                & $\Delta B_{z,NS}$ &      & 0.00  & 0.53 &      & 0.00 & 4.73 &      & 0.00   & 0.28  \\
                & $\Delta B_{x,EW}$ &      &       &      &      &      &      &      & 0.12   & 0.52  \\
                & $\Delta B_{y,EW}$ &      &       &      &      &      &      &      & 0.01   & 1.10  \\
                & $\Delta B_{z,EW}$ &      &       &      &      &      &      &      & 0.01   & 0.53  \\   
\hline 
\end{tabular}
\caption{GVO model rms misfit statistics between GVO estimates and the contributing data for the global grid of 300 GVO's during CHAMP, CryoSat-2 and \textit{Swarm}. Here $\sum$ and $\Delta$ represent data sums and data differences, respectively, split into the North-South (NS) and East-West (EW) components.}
\label{table:2}}
\end{table}
%******************************************

\begin{figure*}[!h]
\centerline{\includegraphics[angle=0, width=\textwidth]{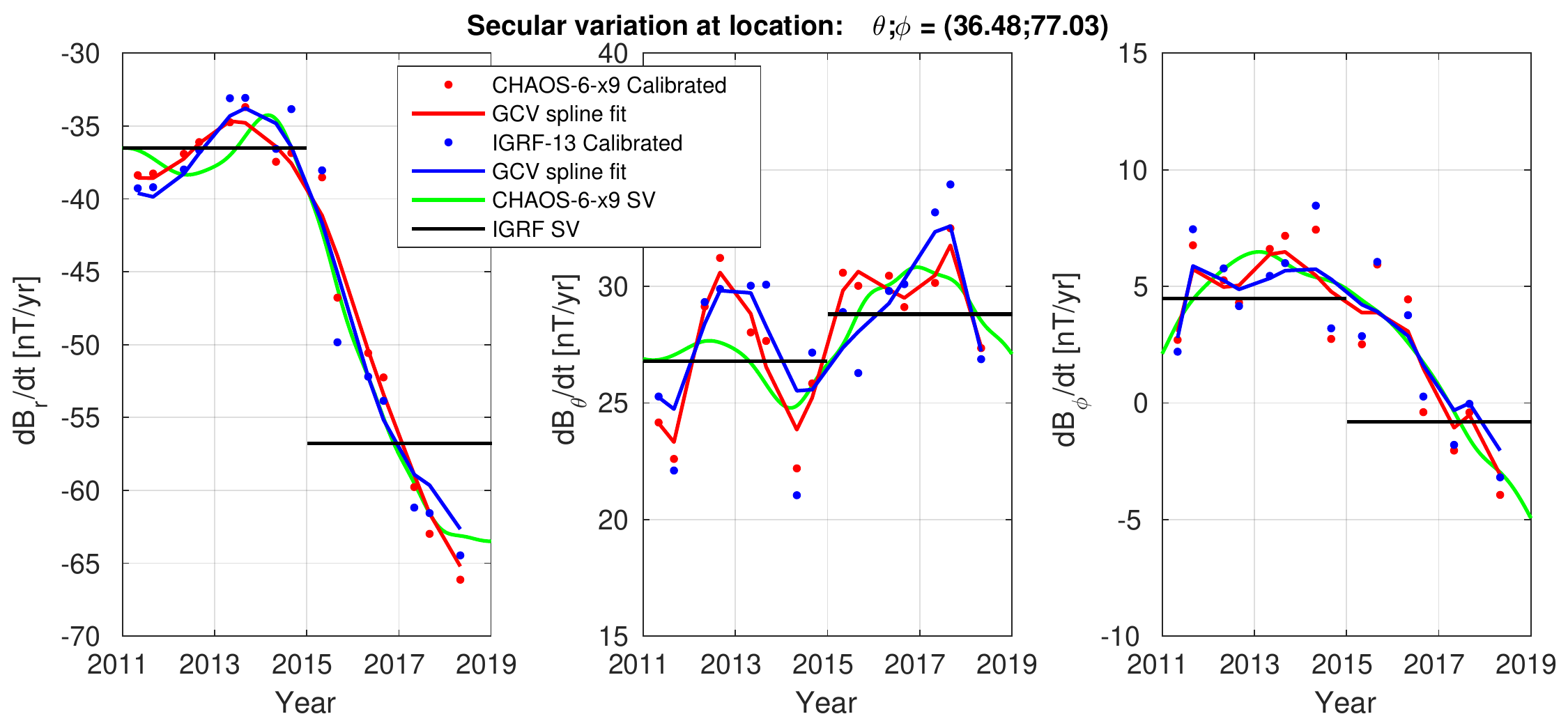}}
\vspace{0.0cm}
\centerline{\includegraphics[angle=0, width=\textwidth]{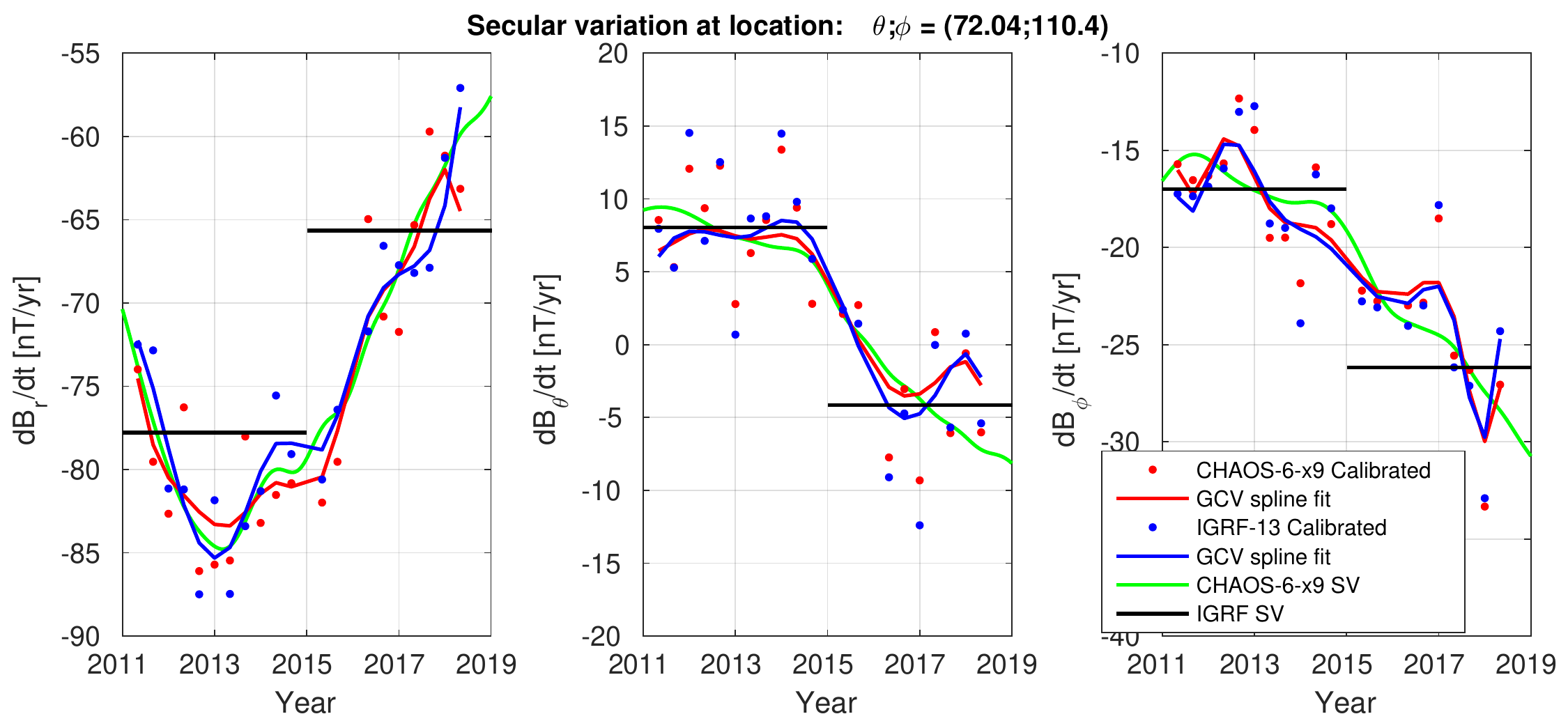}}
\vspace{0.0cm}
\centerline{\includegraphics[angle=0, width=\textwidth]{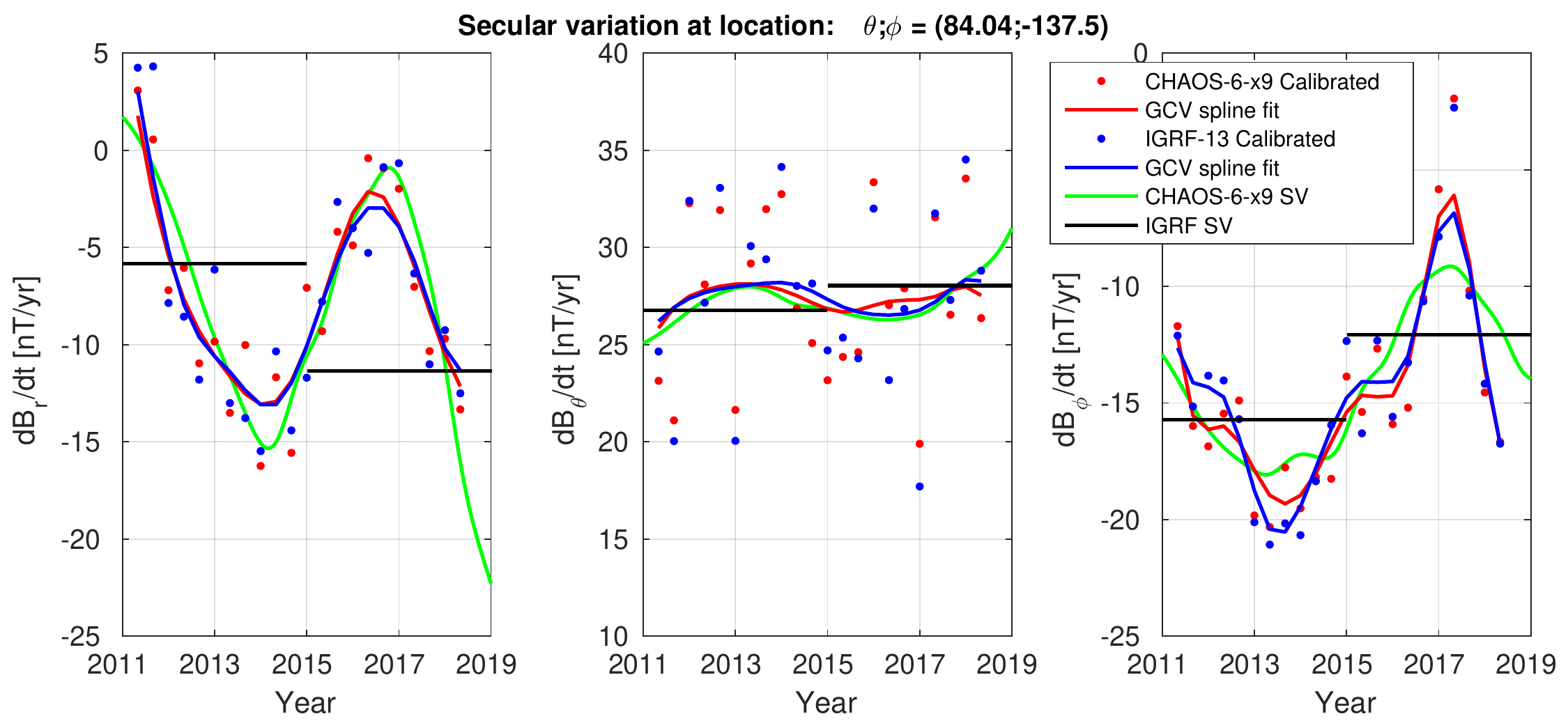}}
\caption{Time series of annual differences of the spherical field components of CryoSat-2 GVO time series derived using CHAOS-6-x9 (red) and IGRF-13 (blue) calibration along with GCV spline fits at the \replaced{three}{two} example grid locations \replaced{: 26 (top), 100 (center) and 140 (bottom)}{100 (top) and 140 (bottom)}. Also shown are \added{IGRF-13 (black) and} CHAOS6-x9 predictions (green).}
\label{Fig:3aa}
\end{figure*}

The CryoSat-2 magnetometer data have been cleaned from known platform signals and calibrated as described by \cite{Olsen_etal_2020}. This calibration relies on computing residuals with respect to a reference magnetic field which was taken from the CHAOS-6-x9 field model \citep{Finlay_etal_2016a}. Here, we carried out an experiment to verify that the GVO secular variation signals obtain from CryoSat-2 data are independent of the main field model used for data calibration. To do this we computed CryoSat-2 GVO estimates using two different datasets, the first being the official dataset calibrated using the CHAOS-6-x9 field model and the second a test version of the CryoSat-2 data calibrated instead using IGRF-13 \citep{Alken_etal_2020}. Having estimated global grids of GVO series for each dataset, to each series we fit cubic smoothing splines, with a knot spacing at every 4 month, and with the smoothing parameter determined using a GCV (generalized cross-validation) approach \citep{Green_Silverman_1993}. Figure \ref{Fig:3aa} presents SV series for the three field components at \replaced{three}{two} example GVOs, with  colatitude/longitude \added{$(36^{\circ},77^{\circ})$ (top),} $(72^{\circ},110^{\circ})$ \replaced{(Center)}{(top)} and  $(84^{\circ},-137^{\circ})$ (bottom). The SV time series derived from the CHAOS-6x9 calibrated dataset are shown with red dots (spline fit in the red line) and similar series from the IGRF-13 calibrated dataset are shown with blue dots (spline fit in the blue line). \added{SV model predictions from the CHAOS-6-x9 model up to SH degree 16 (in green), and IGRF-13 (in black), are also shown for reference.} \deleted{CHAOS-6-x9 model predictions (up to SH degree 16) and  are also shown for reference as the green line.} \added{There are clearly differences in the GVO SV estimates derived using the CHAOS-6x9 and IGRF-13 calibrated data. Since the CryoSat-2 calibration relies on computing residuals with respect to a reference field model (here either CHAOS or IGRF), we do expect such differences, especially since the IGRF model assumes a crude piecewise constant SV field.} \added{Interestingly, the IGRF calibrated SV are an rms of $1.2$nT/yr for the horizontal components, and $3.9$nT/yr for the radial component, closer to CHAOS-6x9 than IGRF-13, indicating that the Cryosat-2 data does possess an SV signal regardless of the model used to calibrate them.}

We find that the CHAOS-6x9 and IGRF-13 calibrated CryoSat-2 GVO SV series show similar sub-decadal changes, neither of which match those seen in CHAOS-6x9.  The similarity of the IGRF and CHAOS calibrated SV series give us confidence that the sub-decadal secular variation seen in CryoSat-2 data is independent of the field model used in its calibration. \added{In addition, we note a strong acceleration in the radial SV component from 2014 to 2018 at the GVO (top plot) located in the Northwest Siberia, a change which is only partially observed in the IGRF-13 model.}

%******************************************
%****************************************** 
In Table \ref{table:1a} we report the mean rms differences between the GVO SV estimates and GCV spline fits, separated by component ($B_r,B_\theta,B_\phi$ and also the intensity $F$) and by polar and non-polar regions. These numbers give an indication of the scatter in the SV datasets and allow the quality of the GVO SV series obtained from CHAMP, \textit{Swarm} and CryoSat-2 (both CHAOS-6x9 and IGRF-13 calibrated versions) to be compared.  Similar results are seen for the CHAOS-6-x9 and IGRF-13 model calibrated GVO SV series, with rms differences between GCV fits and intensity SV data, taking the over all GVOs, of $\replaced{3.5}{3.57}$~nT/yr and  $\replaced{3.4}{3.45}$~nT/yr, respectively. As expected, similar numbers for the CHAMP and especially the \textit{Swarm} derived GVO's, are much smaller being $\replaced{1.8}{1.82}$~nT/yr and $\replaced{0.9}{0.92}$~nT/yr, respectively.  This indicates the lower scatter in the CHAMP and \textit{Swarm} GVO series.

%******************************************
\begin{table}[!h]
\centering
\centerline{\begin{tabularx}{0.7\linewidth}{l | c c c}
\hline
                 & \textit{All}  & \textit{Polar} & \textit{non-Polar} \\
\hline                             
CryoSat-2 CHAOS-6-x9 calibrated                       &          &          &     \\ 
% mean $\sigma_r$         &     3.47 &     4.10 &     3.25\\ 
% mean $\sigma_{\theta}$  &     3.87 &     5.07 &     3.45\\ 
% mean $\sigma_{\phi}$    &     2.92 &     4.90 &     2.23\\
% mean $\sigma_{F}$       &     3.57 &     4.17 &     3.36\\
 mean $\sigma_r$        &     3.38 &     3.98 &     3.17 \\  
 mean $\sigma_{\theta}$ &     3.76 &     4.92 &     3.36\\  
 mean $\sigma_{\phi}$   &     2.85 &     4.77 &     2.17\\  
 mean $\sigma_{F}$      &     3.48 &     4.04 &     3.28\\ 
%\hline  
CryoSat-2 IGRF-13 calibrated                          &          &          &     \\ 
% mean $\sigma_r$         &     3.47 &     3.89 &     3.32\\ 
% mean $\sigma_{\theta}$  &     3.65 &     4.34 &     3.41\\ 
% mean $\sigma_{\phi}$    &     2.84 &     4.77 &     2.16\\ 
% mean $\sigma_{F}$       &     3.45 &     4.02 &     3.25\\ 
 mean $\sigma_r$        &     3.38 &     3.78 &     3.24\\ 
 mean $\sigma_{\theta}$ &     3.54 &     4.20 &     3.31\\  
 mean $\sigma_{\phi}$   &     2.77 &     4.64 &     2.11\\ 
 mean $\sigma_{F}$      &     3.36 &     3.91 &     3.17\\  
 \hline  
CHAMP                            &          &          &     \\ 
% mean $\sigma_r$         &     2.54 &     2.83 &     2.44\\ 
% mean $\sigma_{\theta}$  &     2.18 &     3.67 &     1.65\\ 
% mean $\sigma_{\phi}$    &     1.83 &     2.83 &     1.48\\ 
% mean $\sigma_{F}$       &     1.82 &     2.86 &     1.46\\ 
 mean $\sigma_r$         &     2.49 &     2.76 &     2.40 \\ 
 mean $\sigma_{\theta}$  &     2.14 &     3.61 &     1.62\\ 
 mean $\sigma_{\phi}$    &     1.80 &     2.76 &     1.46 \\ 
 mean $\sigma_{F}$       &     1.78 &     2.79 &     1.43 \\ 
 \hline  
\textit{Swarm}                   &          &          &     \\ 
% mean $\sigma_r$         &     0.94 &     1.38 &     0.78\\ 
% mean $\sigma_{\theta}$  &     1.10 &     2.02 &     0.77\\ 
% mean $\sigma_{\phi}$    &     1.63 &     2.13 &     1.45\\ 
% mean $\sigma_{F}$       &     0.92 &     1.38 &     0.76\\ 
 mean $\sigma_r$         &     0.91 &     1.33 &     0.76 \\ 
 mean $\sigma_{\theta}$  &     1.06 &     1.95 &     0.75 \\ 
 mean $\sigma_{\phi}$    &     1.58 &     2.06 &     1.41\\ 
 mean $\sigma_{F}$       &     0.89 &     1.34 &     0.73\\ 
 \hline 
\end{tabularx}}
\caption{Mean of the rms differences (in nT/yr) between GVO SV series and GCV cubic spline fits.  Results are shown for both the CHAOS-6-x9 and IGRF-13 calibrated CryoSat-2 datasets, and for GVO SV series derived using \textit{Swarm} and CHAMP data.}
\label{table:1a}
\end{table} 
\begin{figure*}[!htbp]
\centerline{\includegraphics[angle=270, width=0.98\textwidth]{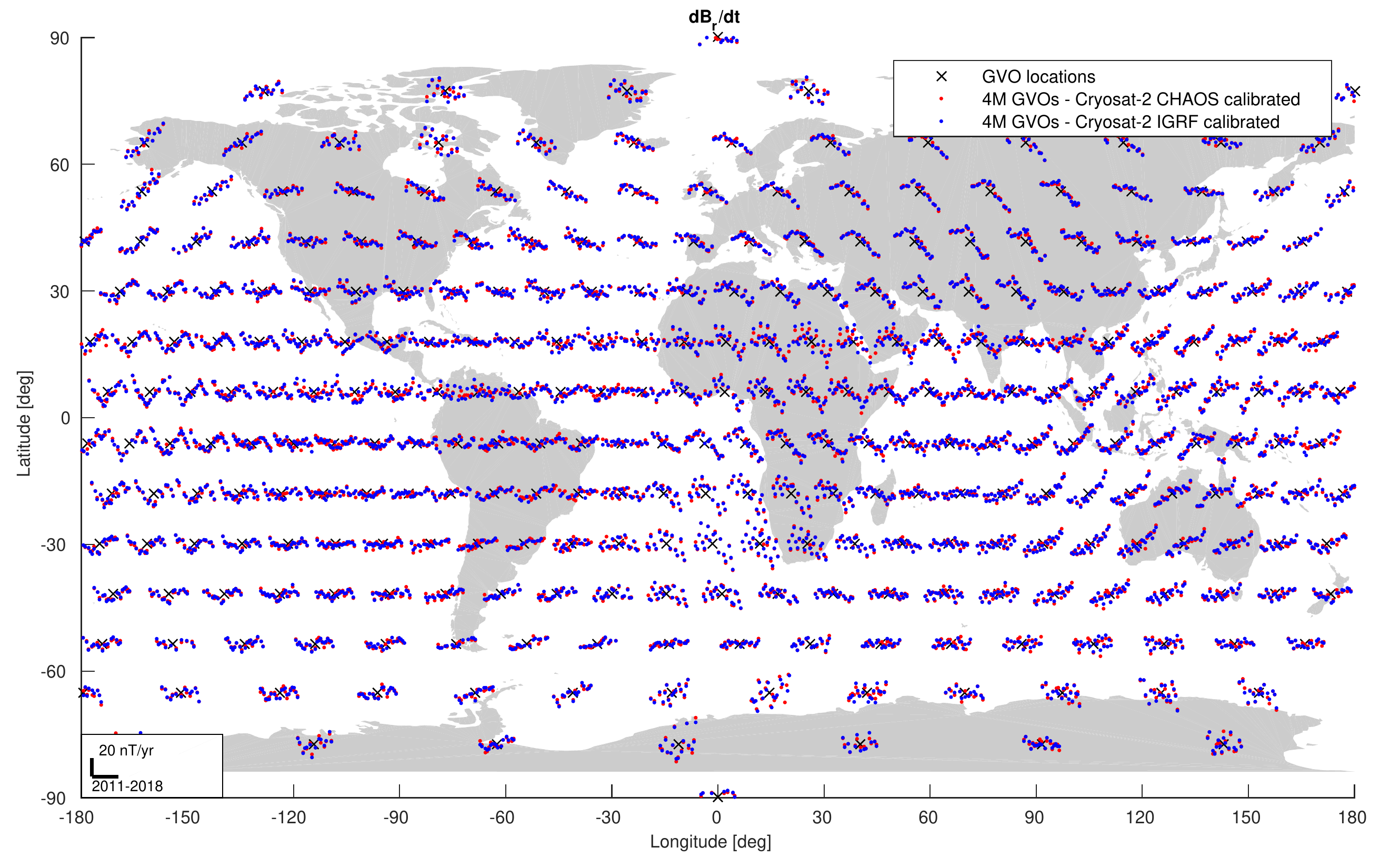}}
\caption{Time series of annual differences of the radial field component for GVO's derive from CryoSat-2 data calibrated using CHAOS-6-x9 (red dots) and IGRF-13 (blue dots), and relocated to a common altitude of 700~km. GVO locations are marked with a black cross.}
\label{Fig:3a}
\end{figure*}
%*****************************************

 Figure \ref{Fig:3a} goes beyond detailed comparisons at a few example locations and presents a global map of CryoSat-2 GVO time series of the radial field SV, showing both the official CHAOS-6x9 calibrated dataset (red dots) and the IGRF-13 calibrated test dataset (blue dots). The scale is shown in the bottom left corner, with the y-axis  being 20\,nT/yr and the x-axis going from 2011 to 2018. \replaced{Visual inspection of Figure \ref{Fig:3a} seems to support the results of Table \ref{table:1a}, and shows similar scatter levels with rms differences of less than $1.0$nT/yr in all three components of the SV series derived from the CHAOS-6-x9 and IGRF-13 calibrations.}{Visual inspection of Figure~\ref{Fig:3a} confirms that both the sub-decadal trends and the scatter in the SV series are comparable for the CHAOS-6-x9 and IGRF-13 calibrations, respectively.}
 
\begin{figure*}[!h]
\centerline{\includegraphics[angle=0, width=\textwidth]{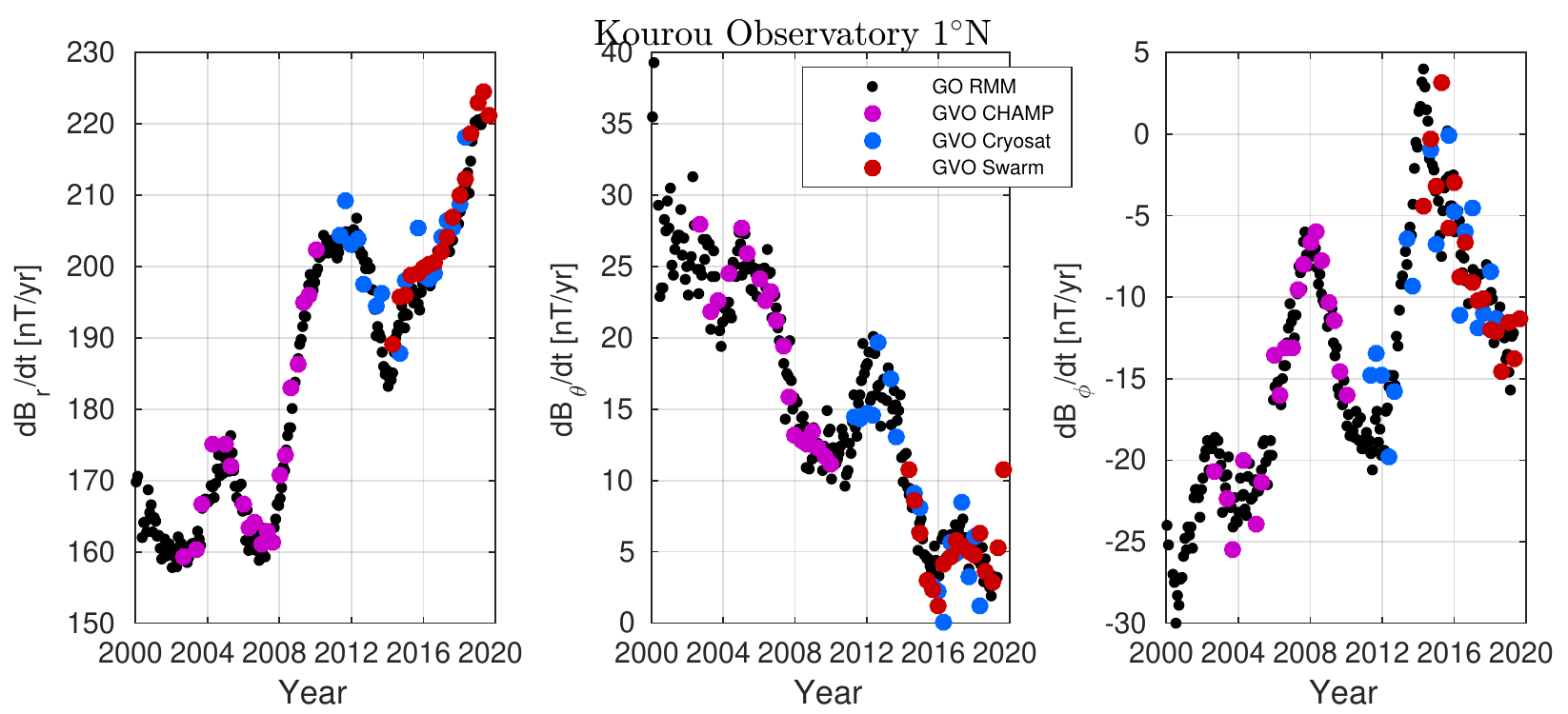}}
\vspace{0.0cm}
\centerline{\includegraphics[angle=0, width=\textwidth]{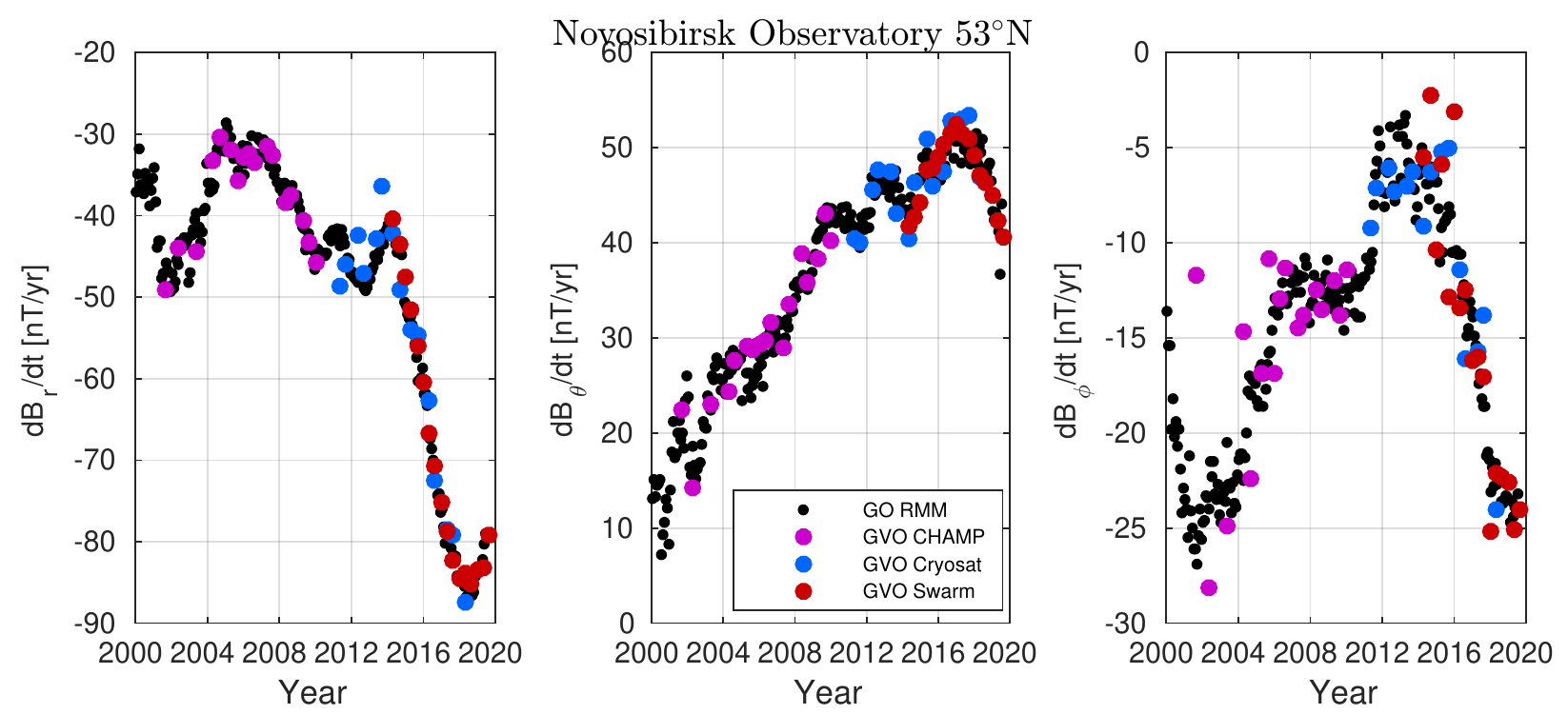}}
\vspace{0.0cm}
\centerline{\includegraphics[angle=0, width=\textwidth]{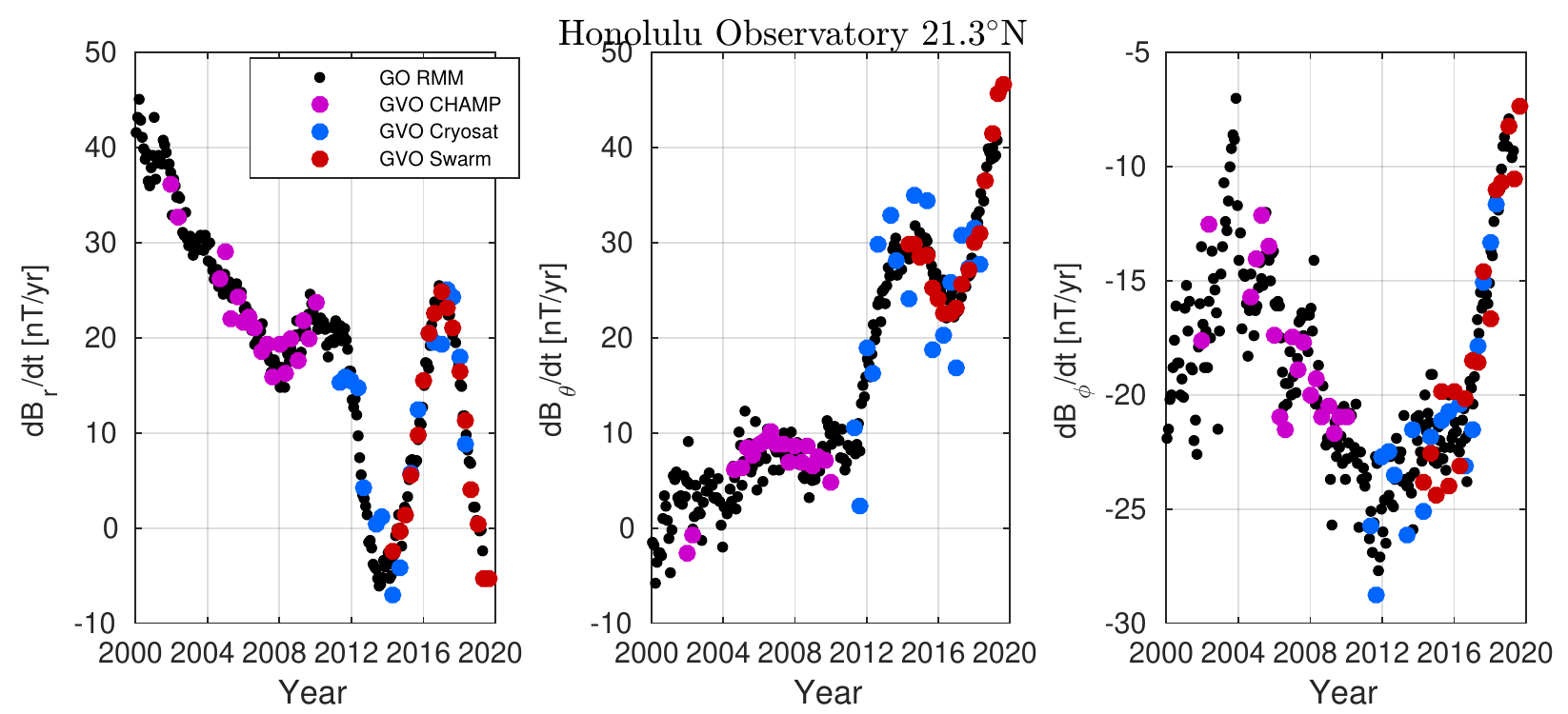}}
\caption{SV times series obtained from annual differences of revised monthly means (black dots), \citep{Olsen_etal_2014}, and 4-monthly Core Field GVOs derived from CHAMP (purple dots), CryoSat-2 (blue dots) and \textit{Swarm} (red dots) data.  Note the GVO estimates have been mapped from satellite altitude to ground in order to aid the comparison.}
\label{Fig:3}
\end{figure*}

With the availability of CryoSat-2 magnetic data, it is now possible to use GVOs to study secular variation globally over the past twenty years. In order to illustrate the quality of information they can provide, in Figure \ref{Fig:3} we present comparisons of GVO SV series, from CHAMP, CryoSat-2 and \textit{Swarm},  with Revised Monthly Means (rmm) \citep{Olsen_etal_2014} from high quality ground observatory SV time series from Kourou  in South America (top plots), from Novosibirsk observatory in Siberia (middle plots) and from Honolulu ground observatory in the central Pacific (bottom plots). Each plot shows the spherical polar components of the annual differences of revised monthly means (black dots) computed from the ground observatory data \citep{Olsen_etal_2014}, and GVO time series derived from CHAMP (purple dots), CHAOS-6x9 calibrated CryoSat-2 (blue dots) and \textit{Swarm} (red dots) data relocated to ground level\replaced{. Here the GVO time series for each satellite mission have been mapped to a common altitude of 700km by subtracting the field difference as given by the CHAOS-7.2 model for SH degrees 1-20, between 700km altitude and GVO altitudes which are close to the mean orbital altitude for each mission.}{using the CHAOS-7.2.} At all three locations the GVOs derived from CryoSat-2 data have more scatter -- this is particularly noticeable in the $\theta$- and $\phi$-components for the examples shown, whereas the scatter in the radial component is closer to the level seen in the CHAMP and \textit{Swarm} derived GVOs. Table \ref{table:1a} indicates that at non-polar latitudes the scatter in the  $r$- and $\theta$-components is generally similar, \replaced{3.4}{3.47}\,nT/yr compared to \replaced{3.8}{3.87}\,nT/yr. Notice that the scatter in the ground observatory rmm's is also enhanced in the horizontal components. The good agreement between the independently estimated CryoSat-2 and \textit{Swarm} derived GVO's at overlapping epochs from 2014 to 2018, is particularly evident in the $r$- and $\phi$-components. SV variations are coherent in both phase and amplitude between the CryoSat-2 GVO time series and the ground observatory records, thus confirming that the CryoSat-2 GVO's are able to track the same field changes as observed by ground observatory records on timescales of 1~year and longer.  

%******************************************
%******************************************

\begin{figure*}[!htbp]
\centerline{\includegraphics[angle=270, width=0.98\textwidth]{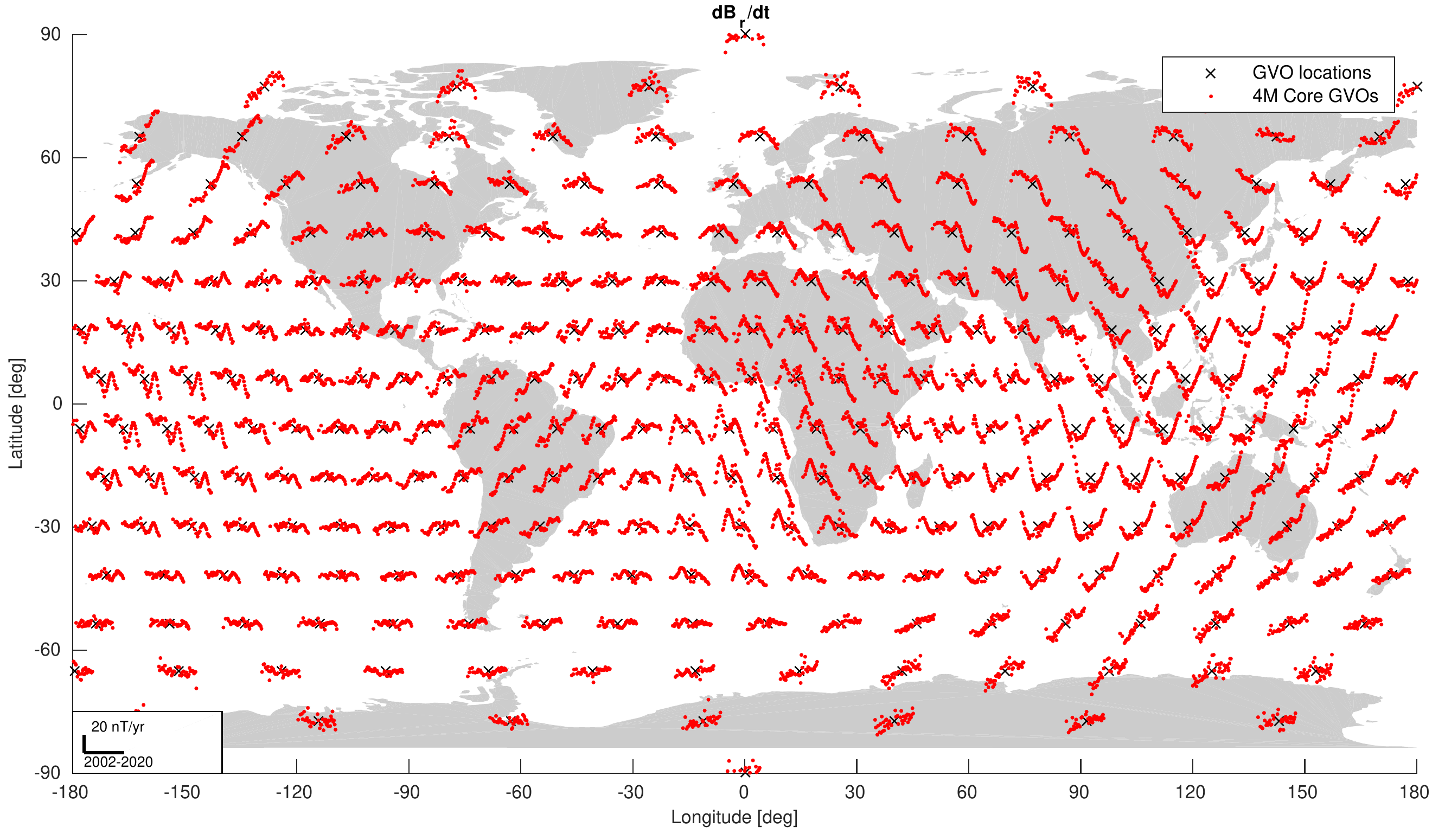}}
\caption{Composite GVO SV time series for the radial field component, computed using annual differences of 4 monthly Core Field GVOs (red dots) which have been mapped to a common altitude of 700km. GVO locations are marked with a black cross.}
\label{Fig:4}
\end{figure*}

Next, in Figure \ref{Fig:4} we presents a global map of the SV of the radial field component over the past 18 years. Here, to ease visualization, GVO time series from CHAMP (covering 2002-2010), CHAOS-6x9 calibrated CryoSat-2 (covering 2010-2014) and \textit{Swarm} (covering 2014-2020) have been mapped to a common altitude of 700~km, again using the CHAOS-7.2 field model, and combined into one composite time series. This allows for the investigation of global patterns of sub-decadal SV. We find that regions at low latitudes display strong sub-decadal variations, however, not simultaneously at all longitudes. For instance, we observe a change of slope in the radial SV field occurring over the south Atlantic region around 2007 and again in 2014, over Indonesia around 2014, and in the Pacific region centred in 2017. Some of these variations are characterised by distinct "$\Lambda$" and "$V$"-shaped behaviour occurring over time spans of 5-10 years and locally confined to specific regions, but otherwise reminiscence of often discussed geomagnetic jerks \citep{Mandea_etal_2010}. The availability of CryoSat-2 magnetic field data clearly plays \replaced{a key role in permitting a continuous coverage from satellite-based time series without a gap between 2010 and 2013, thus}{an important role in these continuous satellite-based time series,} allowing the study of global patterns in the time-varying core field secular variation over the past twenty years.
%******************************************
%******************************************
\newpage

\section{Application II: SOLA}
\label{sec:SOLA}

\subsection{SOLA Method for local estimation of CMB radial field SV}
\label{sec:SOLA_method}
We now move on to investigate the behaviour of the core field not at satellite altitude but down at the core-mantle boundary (CMB), on the edge of the region where it originates. To do this we use the Green’s functions of the Neumann boundary value problem that links the magnetic field at satellite altitude to the radial field at the CMB. Following \cite{Hammer_Finlay_2019}, a localized estimate, $\widehat{B}_r$, of the radial magnetic field at a target location and time, $(\mathbf{r}_0,t_0)$, at the CMB can be computed as a localized spatial average around the target location time-averaged over a specified interval. Because the CMB radial magnetic field is linearly related to the spherical polar components of the vector field at satellite altitude, we can write  $\widehat{B}_r$ as a weighted linear combination of the satellite magnetic measurements with weights $q_n$ \citep{Backus_Gilbert_1968,Backus_Gilbert_1970,Hammer_Finlay_2019} 

\begin{equation}
\widehat{B}_r(\mathbf{r}_0,t_0)  = \sum_n^N q_n(\mathbf{r}_0,t_0) \, d_n(\mathbf{r}_n,t_n) \label{eq:SOLA_est}
\end{equation}

where $d_n$ are satellite magnetic measurements $(n=1,...,N)$ within a specified time window and $q_n$ are weighting coefficients to be determined. Here the data $d_{n}$, at positions $\mathbf{r}_n$ and times $t_n$, are taken from dataset \#2 and for simplicity we consider \replaced{using}{use} only observations of the radial component of the field. Corrections for the lithospheric field for SH degrees $n \in [14,185]$ as given by the LCS-1 model \citep{Olsen_etal_2017}, for the magnetospheric and associated induced fields as given by the CHAOS-7.2 model \citep{Finlay_etal_2020}, and for the ionospheric and associated induced fields as given by the CIY4 model, \citep{Sabaka_etal_2018} are removed from the observations in a pre-processing step. The radial magnetic field measurements, $d_{n}(\mathbf{r}_n,t_n)$, are then related to the radial magnetic field $B_r(\mathbf{r}',t_n)$, integrated over the CMB, by \citep{Gubbins_Roberts_1983}

\begin{equation}
d_{n}(\mathbf{r}_n,t_n) = \oint_{S'} G_{r}(\mathbf{r}_n, \mathbf{r}')B_r(\mathbf{r}',t_n) dS'  \label{eq:SOLA_data_int1}
\end{equation}

where the surface element is $dS'=\mathrm{sin}\theta'd\theta'd\phi'$. The data kernel $G_{r}(\mathbf{r}_n, \mathbf{r}')$ is the radial derivative with respect to $\mathbf{r}$, of the Green's functions for the exterior Neumann boundary value problem \citep[e.g.][]{Gubbins_Roberts_1983, Barton_1989}

\begin{equation}
G_{r}      = \frac{1}{4\pi}\frac{h_n^2(1-h_n^2)}{f_n^3} \label{eq:SOLA_data_kernel}
\end{equation}
  
where $h_n=r'/r_n$ and $r'$ is the CMB radius, $f_n=R_n/r_n$ where $R_n=\sqrt{r_n^2+r'^2-2r_nr'\zeta_n}$ and $\zeta_n = \mathrm{cos}\, \gamma_n=\mathrm{cos}\,\theta_n \, \mathrm{cos}\,\theta'+\mathrm{sin}\,\theta_n \mathrm{sin}\,\theta'\, \mathrm{cos(\phi_n-\phi')}$, where $\gamma_n$ is the angular distance between a measurement at position $(\theta_n,\phi_n)$ and a position on the CMB $(\theta',\phi')$. The data kernel describes how a particular measurement samples the CMB radial field; radial magnetic measurements sample the CMB radial field most strongly directly below the measurement position. Regarding the time-dependence, we use a first order Taylor expansion around a reference time $t_0$, such that

\begin{equation}
d_{n}(\mathbf{r}_n,t_n) \approx \oint_{S'} G_{r}(\mathbf{r}_n,\mathbf{r}') \left[ B_r(\mathbf{r}',t_0)+\dot{B_r}(\mathbf{r}',t_0)\Delta t_n\right] dS' \label{eq:SOLA_data_int2}
\end{equation}
 
The time difference, $\Delta t_n = t_n-t_0$, is computed with respect to the target time, $t_0$. Inserting eq.\eqref{eq:SOLA_data_int2} into eq.\eqref{eq:SOLA_est} we obtain

\begin{equation}
\widehat{B}_r(\mathbf{r}_0,t_0)  =  \oint_{S'} \mathcal{K}(\mathbf{r}_0, t_0, \mathbf{r}') \, B_r(\mathbf{r}',t_0)dS' + \oint_{S'} \mathcal{\dot{K}}(\mathbf{r}_0, t_0, \mathbf{r}') \, \dot{B_r}(\mathbf{r}',t_0) dS' \label{eq:SOLA_est2}
\end{equation}

where $\mathcal{K}(\mathbf{r}_0, t_0, \mathbf{r}')$ and $\mathcal{\dot{K}}(\mathbf{r}_0, t_0, \mathbf{r}')$ are spatial averaging kernels for the CMB field and secular variation respectively, constructed from the weighting coefficients and the data kernels 
\begin{align}
\mathcal{K}(\mathbf {r}_0, t_0, \mathbf{r}') &    = \sum_n^N q_n(\mathbf{r}_0,t_0) \, G_r(\mathbf{r}_n,\mathbf{r}') \label{eq:SOLA_kernel_1}  \\
\mathcal{\dot{K}}(\mathbf{r}_0, t_0, \mathbf{r}')& = \sum_n^N q_n(\mathbf{r}_0,t_0) \, G_r(\mathbf{r}_n,\mathbf{r}')\Delta t_n \label{eq:SOLA_kernel_2}
\end{align}
By varying the weight coefficients, $q_n$, the shape of the averaging kernels change. Notice that time differences $\Delta t_n$, between the measurement times and the target time, are effectively additional temporal weights applied to the kernel $\mathcal{K}$ in order to obtain $\mathcal{\dot{K}}$. 

\begin{figure*}[!h]
\centerline{\includegraphics[angle=0, width=0.7\textwidth]{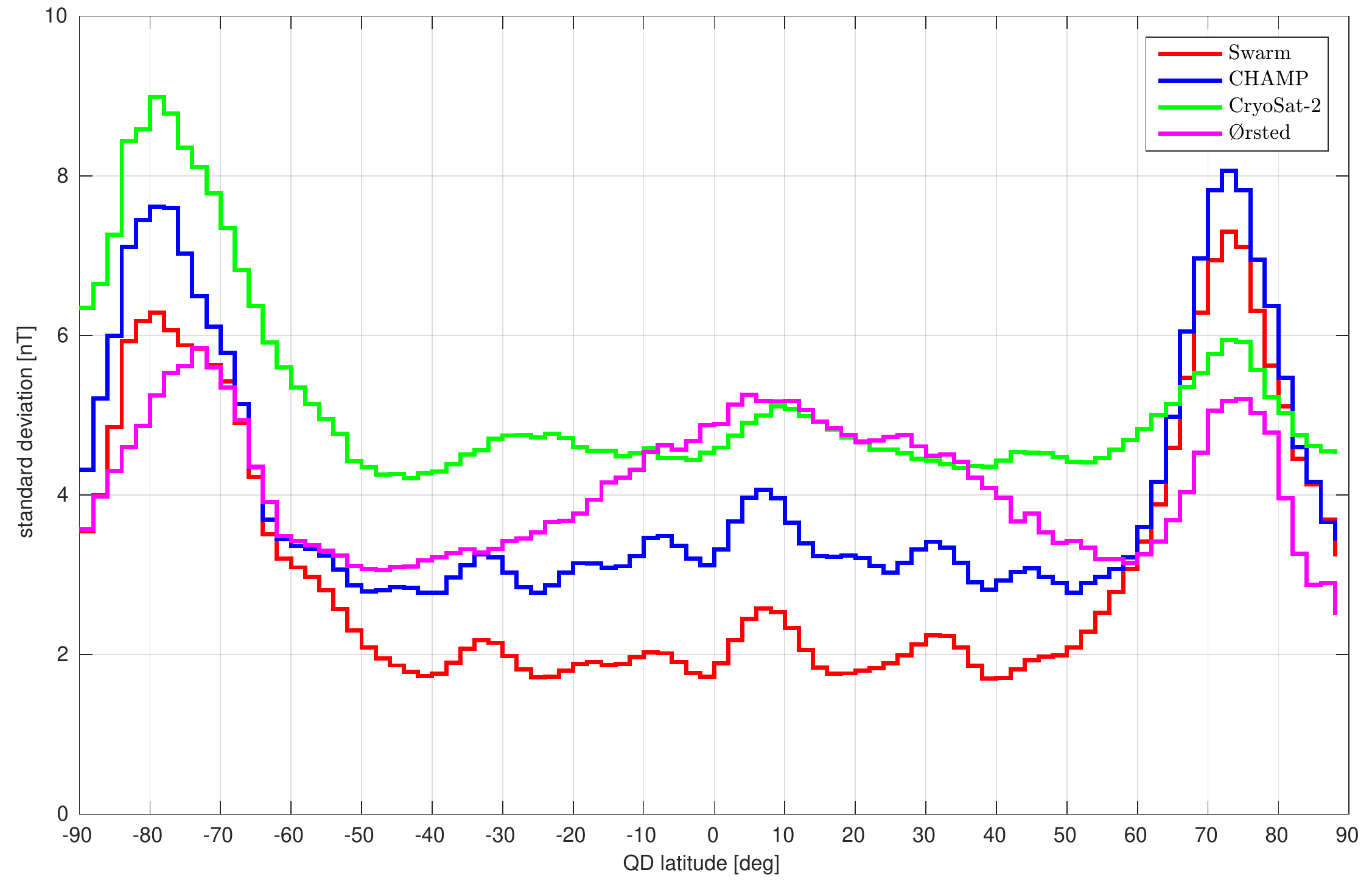}}
\caption{Latitude-dependent standard deviations, $\sigma(\theta_{QD})$, contributing to the data error budget. For the radial field component in $2^{\circ}$ bins (Northern hemisphere having positive QD latitude) for each satellite mission.}
\label{Fig:7}
\end{figure*}

In order to obtain estimates of the secular variation of the radial field on the CMB, at the target location and time $\widehat{\dot{B}}_r(\mathbf{r}_0,t_0)$, we minimize the following objective function

\begin{equation}
\Theta = \oint_{S'} [\mathcal{K}(\mathbf{r}_0, t_0, \mathbf{r}')]^2dS' +\oint_{S'} [\dot{\mathcal{K}}(\mathbf{r}_0, t_0, \mathbf{r}')-\dot{\mathcal{T}}(\mathbf{r}_0, \mathbf{r}')]^2dS' + \lambda^2 \mathbf{q}^T \underline{\underline{\mathbf{E}}} \mathbf{q} \label{eq:SOLA_obj}
\end{equation}

where $\lambda$ is a trade-off parameter (units of $[\mathrm{nT}^{-1}]$), $\mathbf{q}$ is vector of the weighting coefficients, $\underline{\underline{\mathbf{E}}}$ is the data error covariance matrix which we define below and $\dot{\mathcal{T}}$ is an SV target kernel that we choose to be a Fisher distribution on the sphere \citep{Fisher_1953}
\begin{equation}
\dot{\mathcal{T}}(\mathbf{r}_0, \mathbf{r}') =\frac{\kappa}{4 \pi \mathrm{sinh}\kappa} e^{\kappa \cos \, \gamma_0}  \label{eq:Kernel_target2}
\end{equation}
 where $\gamma_0$ is the angular distance on the CMB between the target position $(\theta_0,\phi_0)$ and another position $(\theta',\phi')$. On the basis of tests carried out by \cite{Hammer_Finlay_2019}, we set $\kappa=600$ corresponding to a target kernel width of $15^{\circ}$; this is narrower than can be achieved for $\dot{\mathcal{K}}$ with the available data, but it avoids excessive ringing associated with taking a Dirac delta function as the target kernel. \replaced{When computing SOLA estimates for a given time window, we select a subset ($n=1,...,N$) of the measurements. Using this data subset the}{The} data error covariance matrix $\mathbf{E}$ is defined as follows. Using all available measurements ($m=1,...,M$), for each satellite mission within 2 degree bins of quasi-dipole (QD) latitude \citep{Richmond_1995}, we first derived robust data error variances as a function of QD latitude

\begin{equation}
\sigma^2(\theta_{QD}) =  { \sum\limits_{m=1}^M w_m\left(\epsilon_m-\mu\right)^2 \over \sum\limits_{m=1}^M w_m }\label{eq:sigma_QD}
\end{equation}

where $\epsilon_m$ are residuals with respect to predictions of the CHAOS-7.2 internal field model for SH degrees $n \in [1,13]$, $\mu$ are robust mean residuals within the considered bin and $w_m$ are Huber weights \citep[e.g.,][]{Constable_1988} for the data within each bin. \added{Here we use QD coordinates, as this is appropriate for characterizing processes related to unmodelled ionospheric currents which we consider to be a likely source of contamination, especially at high latitudes.} Figure \ref{Fig:7} presents the resulting QD-latitude-dependent error estimates $\sigma(\theta_{QD})$ for the radial field component used in this study, comparing the values for the {\O}rsted, CHAMP, CryoSat-2 and {\it Swarm} datasets. \added{When computing SOLA estimates for a specified time window of e.g. 2yrs, we select a data subset of dataset \#2. Using this data subset of N measurements, a data error covariance matrix E computed.}
Diagonal elements of \replaced{this}{the} data error covariance matrix $\mathbf{E}$ are finally defined as

\begin{equation}
E_{n,n}=\sigma^2(\theta_{QD}) / w_n \label{eq:sigma_QD_n}
\end{equation}

where $w_n$ are robust (Huber) weights determined a-priori for each datum ($n=1,...,N$), based on their residual to CHAOS-7.2, in order to account for the expected long-tailed error distribution. Off-diagonal elements of $\mathbf{E}$ are set to zero.

In addition to minimizing eq.\eqref{eq:SOLA_obj} we simultaneously impose the following constraint 

\begin{equation}
\oint_{S'}\mathcal{K}(\mathbf{r}_0, t_0, \mathbf{r}') dS' + \oint_{S'} \dot{\mathcal{K}}(\mathbf{r}_0, t_0, \mathbf{r}')dS' = 1 \label{eq:SOLA_norm}
\end{equation}

where the first term is in practise very small when estimating the SV, since it is minimized in the objective function.  This constraint ensures that a valid averaging kernel is obtained.

Discretization of integrals over the CMB was carried out using Lebedev quadrature \citep{Lebedev_Laikov_1999} and the system of equations was solved for the coefficients, $q_n$, using a Lagrange multiplier method, see \cite{Hammer_Finlay_2019} for further details. 

Once SOLA estimates of the CMB radial field SV at the chosen target location and epoch are obtained, by minimizing eq.~\eqref{eq:SOLA_obj} subject eq.~\eqref{eq:SOLA_norm}, we are able to easily appraise them based on (i) their averaging kernel width, which we define as the angular distance between the points at which the averaging kernel first reaches zero amplitude moving away from its maximum value, and (ii) the variance of the SOLA estimate which we computed as

\begin{equation}
\hat{\sigma}^2(\mathbf{r}_0,t_0) = \mathbf{q}^T \underline{\underline{\mathbf{E}}}\mathbf{q} \label{eq:SOLA_var}
\end{equation}

By changing the parameter $\lambda$ of eq.\eqref{eq:SOLA_obj}, a range of solutions can be computed which describes a trade-off between having an averaging kernel width as small as possible and the variance of the estimate being as small as possible \citep{Parker_1977}.  Below we discuss the effect of changing $\lambda$ on our results.

%****************************************
%\newpage
%******************************************

\subsection{Results: SOLA estimates of CMB SV and SA from {\O}rsted, CHAMP, Cryosat-2 and \textit{Swarm} data}
\label{sec:SOLA_results}
\replaced{
We begin by first comparing SOLA estimates for the CMB radial field SV obtained using separate data subsets from the \textit{Swarm} and CryoSat-2 missions respectively. Firstly, \textit{Swarm} and CryoSat-2 datasets are extracted from the main dataset \#2 described in Section \ref{sec:data}, and each covering the same two year time window from 2015.0 to 2017.0}{We begin by comparing SOLA results for the first time derivative (SV) of the radial field at the CMB, obtained using \textit{Swarm} and CryoSat-2 data subsets extracted from dataset \#2 and covering a two year time window from 2015.0 to 2017.0.} \replaced{Next, in order to}{To} obtain data subsets with suitable spatial and temporal coverage, we considered bins surrounding each point in an approximately equal-distance grid at satellite altitude of $\approx2.5^{\circ}$ spacing, based on the partitioning algorithm of \cite{Leopardi_2006}, and randomly sampled one datapoint from each bin, resetting the bins every two months. Data subsets spanning the full two year window from 2015.0 to 2017.0 were produced by accumulating these two monthly globally distributed subsets. The resulting \textit{Swarm} and Cryosat-2 data subsets spanning 2015.0 to 2017.0 consisted of 62469 and 54685 radial field observations respectively.

\begin{figure*}[htbp]
\setcounter{subfigure}{0}
\centerline{
\subfloat[Radial SV from CryoSat\added{-2} data.]{\includegraphics[angle=0, width=0.5\textwidth]{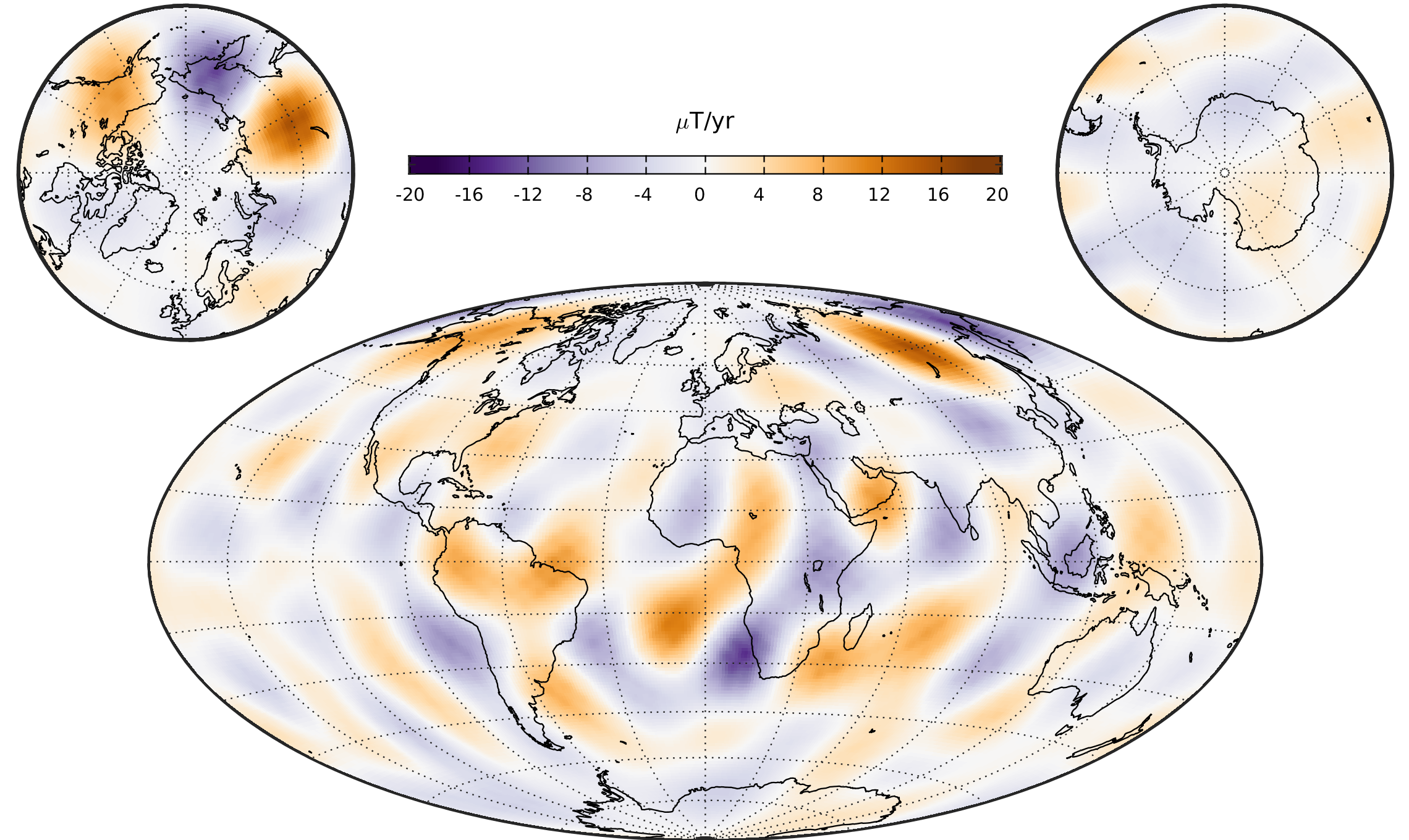}}
\hspace{0.1cm} 
\subfloat[Radial SV from {\it Swarm} data.]{\includegraphics[angle=0, width=0.5\textwidth]{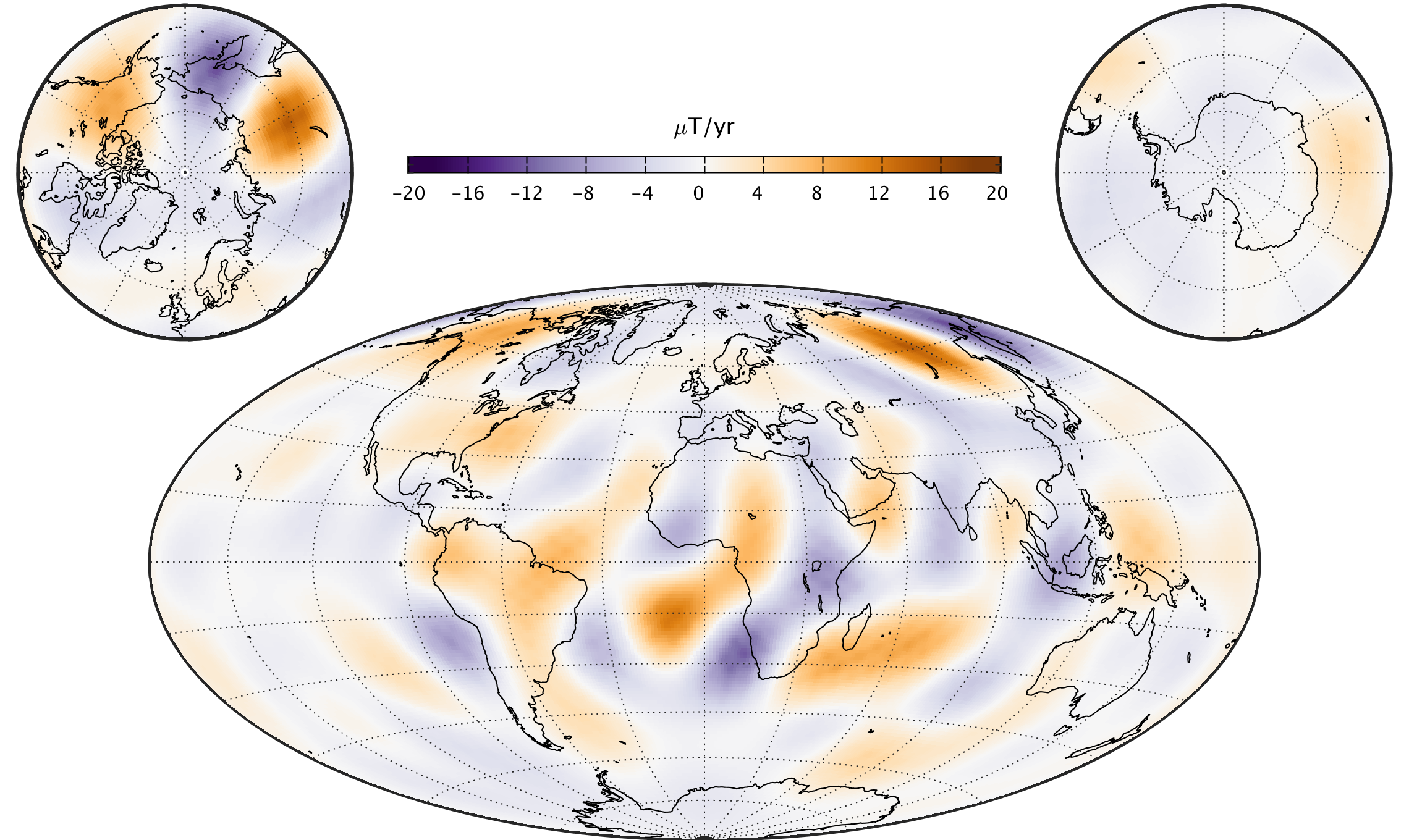}}} 
\centerline{
\subfloat[Error estimates from CryoSat\added{-2} data.]{\includegraphics[angle=0, width=0.5\textwidth]{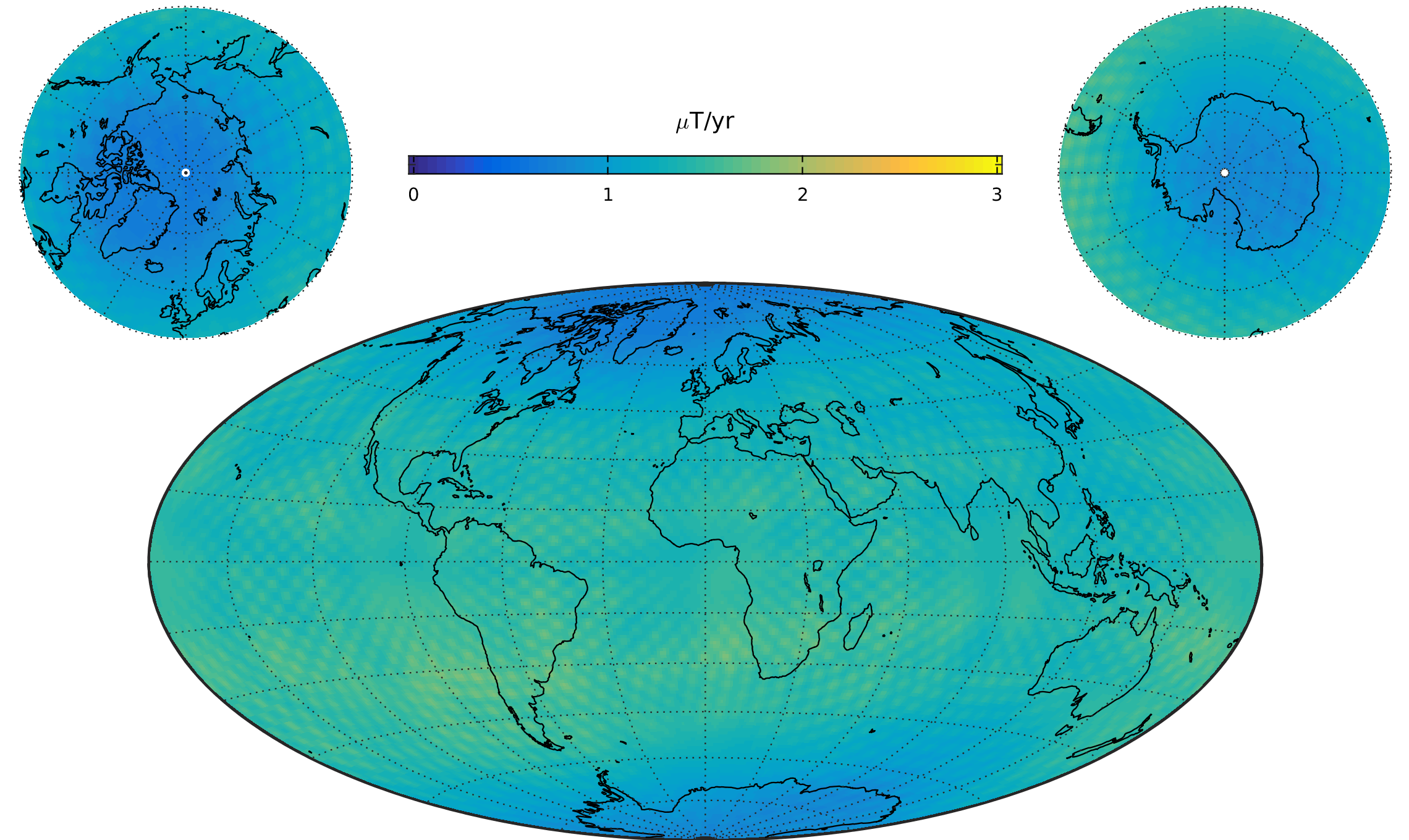}}
\hspace{0.1cm} 
\subfloat[Error estimates from {\it Swarm} data.]{\includegraphics[angle=0, width=0.5\textwidth]{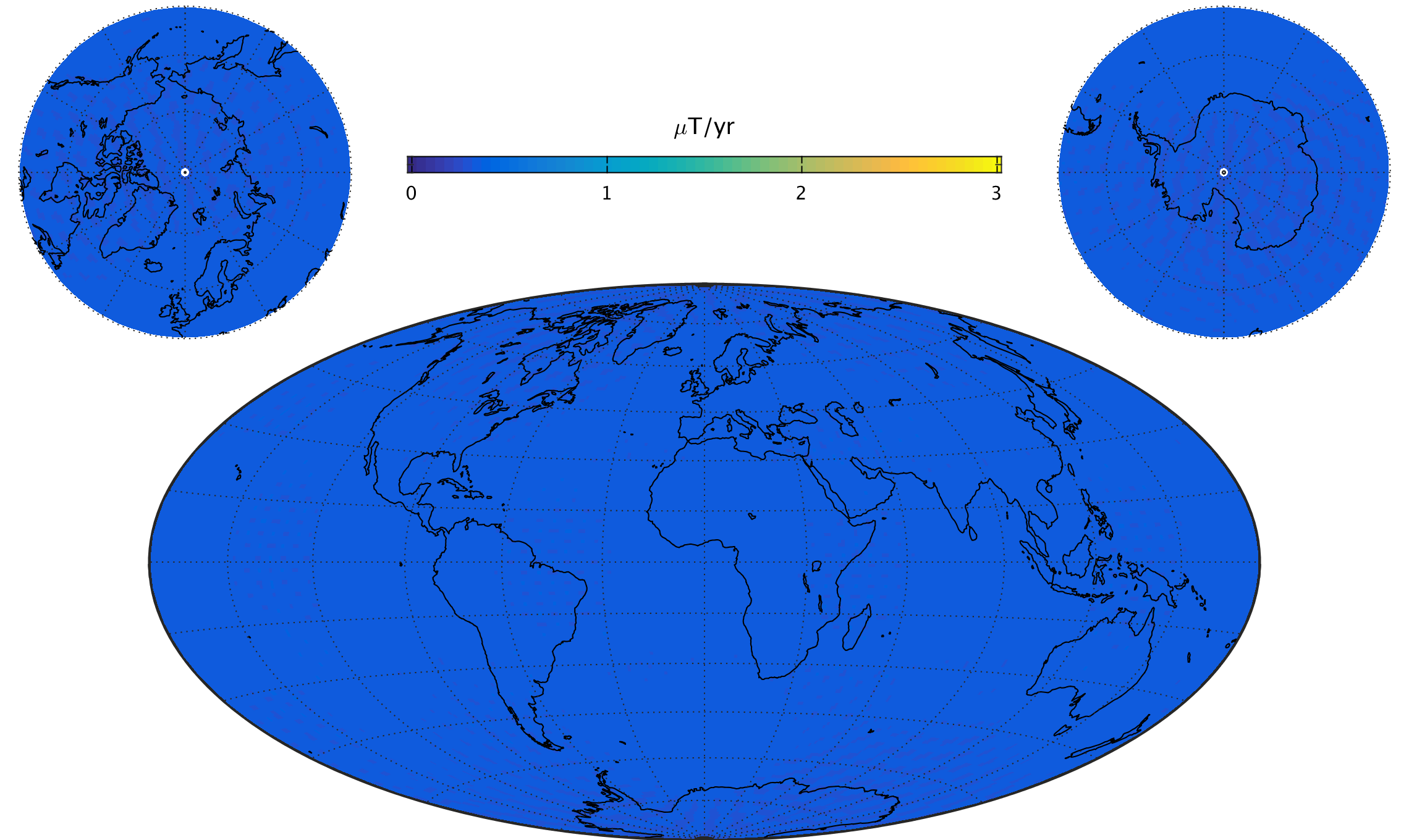}}} 
\centerline{
\subfloat[Averaging kernel widths from CryoSat\added{-2} data.]{\includegraphics[angle=0, width=0.5\textwidth]{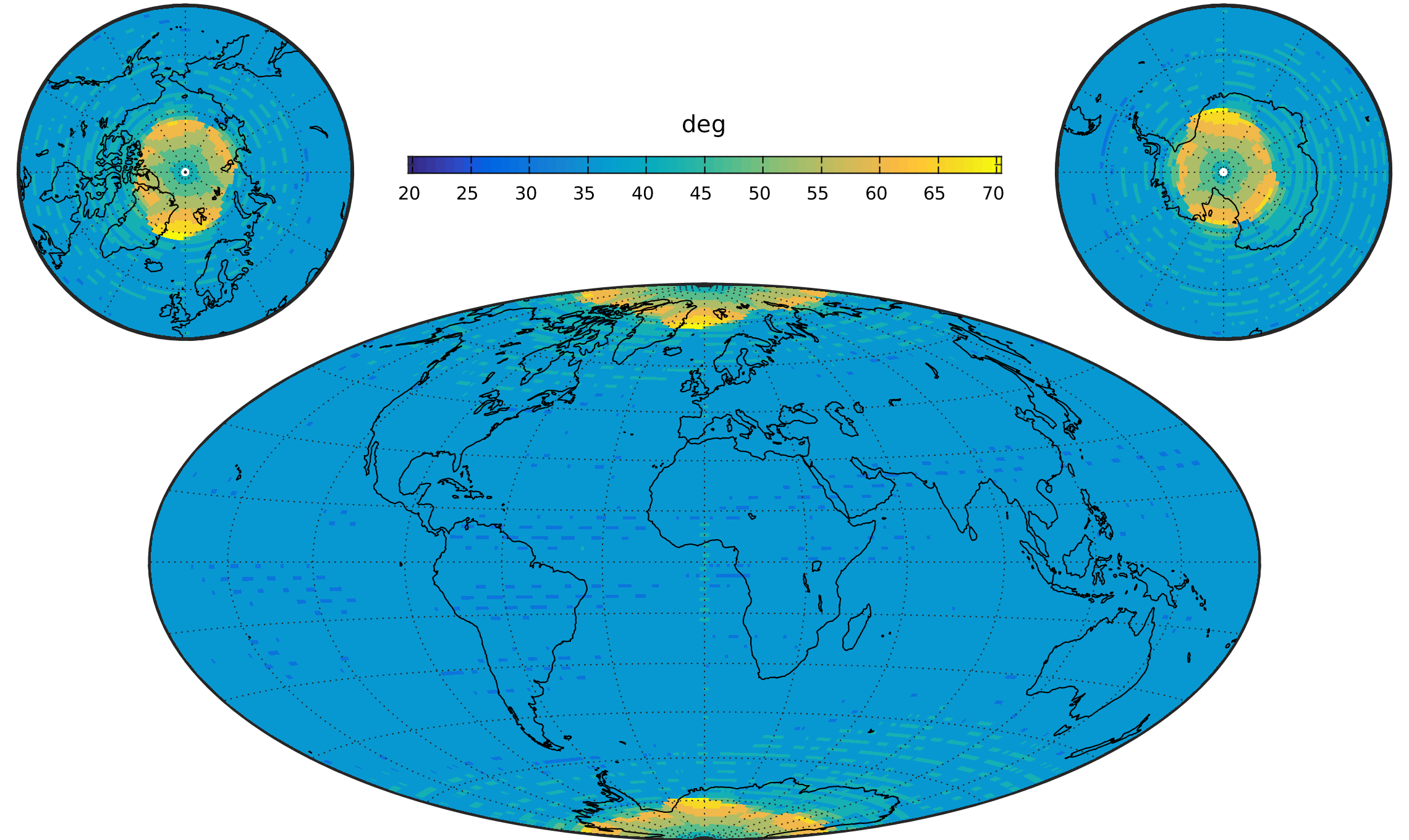}}
\hspace{0.1cm} 
\subfloat[Averaging kernel widths from {\it Swarm} data.]{\includegraphics[angle=0, width=0.5\textwidth]{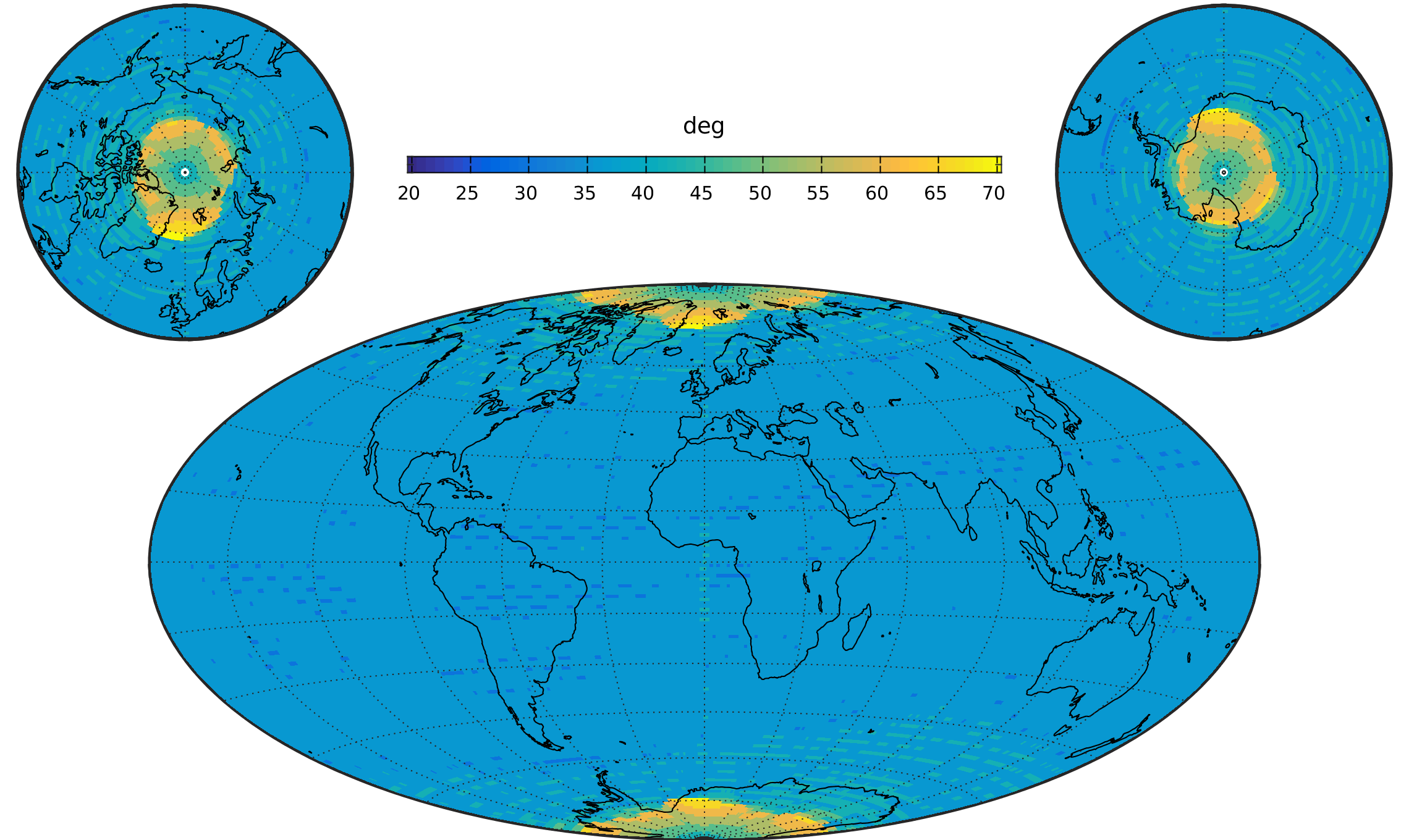}}} 
\caption{SOLA CMB radial field SV estimates for epoch 2016.0, derived using data from 2015.0 to 2017.0 (top plots), associated error estimates (middle plots) and averaging kernel widths (bottom plots). Estimates are derived using CryoSat-2 (left plots) and {\it Swarm} (right plots) data, respectively.}
\label{Fig:9}
\end{figure*}

In Figure \ref{Fig:9} we compare maps collecting SOLA CMB radial field SV estimates centered on epoch 2016.0, derived using the CryoSat-2 and {\it Swarm} data subsets spanning 2015.0 to 2017.0. To ensure that we obtained SOLA estimates of comparable resolution, we first computed SOLA SV estimates using the {\it Swarm} data subset and taking $\lambda=3 \times 10^{-3}\ \mathrm{nT}^{-1}$.  This resulted in well-behaved averaging kernels with widths $\approx42^{\circ}$. Next, we used these averaging kernels as the target kernels in order to derive similar estimates using the {\it Swarm} and CryoSat-2 data subsets, thus effectively seeking {\it Swarm } and CryoSat-2 SV estimates with the same spatial resolution as the target estimates. The maps in the top row of Figure \ref{Fig:9} show the resulting global collections of SOLA SV estimates, obtained on a $1^{\circ}$ grid of target locations at the CMB, based on the CryoSat-2 (left plot) and {\textit Swarm} (right plot) data subsets, respectively.

%******************************************
%******************************************

Maps collecting the formal standard deviations for each SOLA estimate (derived using eq.\ref{eq:SOLA_var}) and their averaging kernel widths are presented in middle and bottom rows of Figure \ref{Fig:9}. The standard deviations of the \textit{Swarm}-based SOLA SV estimates are fairly homogeneous with values of $0.3-0.4$\replaced{$\mu$}{n}T/yr;  those for CryoSat-2 SOLA are somewhat larger, being in the range $\approx0.6-1.8$\replaced{$\mu$}{n}T/yr. We note that the CryoSat-2 error estimates are slightly lower at higher latitudes. The same is true for the {\it Swarm} map, but here the variations in errors estimates are in that case much smaller. Kernel widths in both cases are also fairly homogeneous except at auroral latitudes where distinct behaviour of the kernels are found related to the data error estimates having increased amplitude, as seen in Figure~\ref{Fig:7}. 

As seen from the kernel widths in the bottom row of Figure \ref{Fig:9}, very similar resolution has been obtained (by construction) in the \textit{Swarm} and Cryosat-2 SV maps. The same field features are clearly identified in both maps. For instance, we notice high latitude SV patches in the northern hemisphere which have been associated with a high latitude jet of core flow \citep{Livermore_etal_2017}, and there is increased amplitude of SV over the hemisphere centered on the Atlantic in comparison with the Pacific hemisphere. This first test gives us confidence that the CryoSat-2 measurements can be used to reliably map SV features also at the CMB on a timescale of 2 years, and with a resolution down to $42^{\circ}$ degrees. 

\begin{figure*}[htbp]
\setcounter{subfigure}{0}
\centerline{
\subfloat[Radial SA based on CryoSat\added{-2} data.]{\includegraphics[angle=0,width=0.5\textwidth]{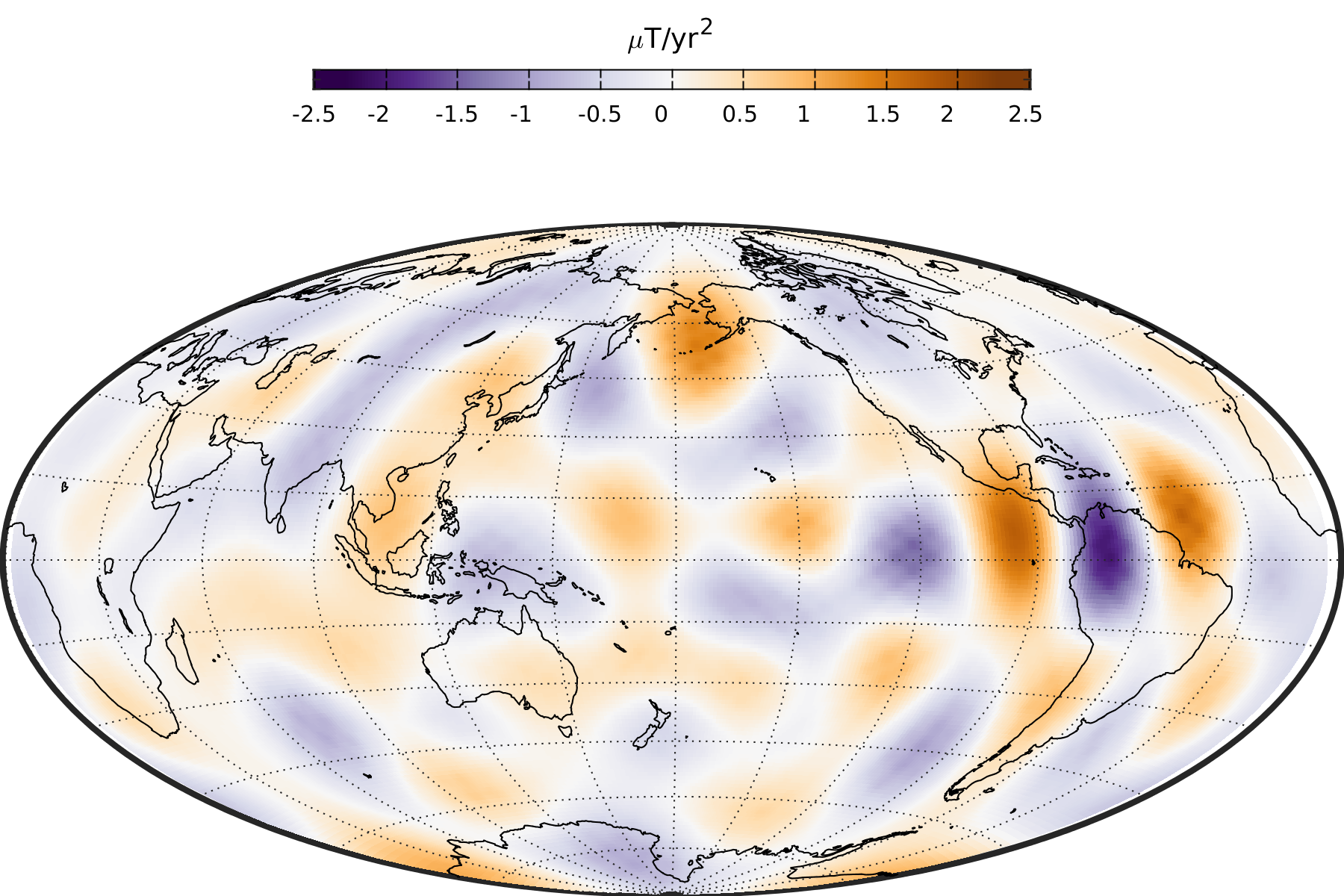}}
\hspace{0.1cm} 
\subfloat[Radial SA based on {\it Swarm} data.]{\includegraphics[angle=0,width=0.5\textwidth]{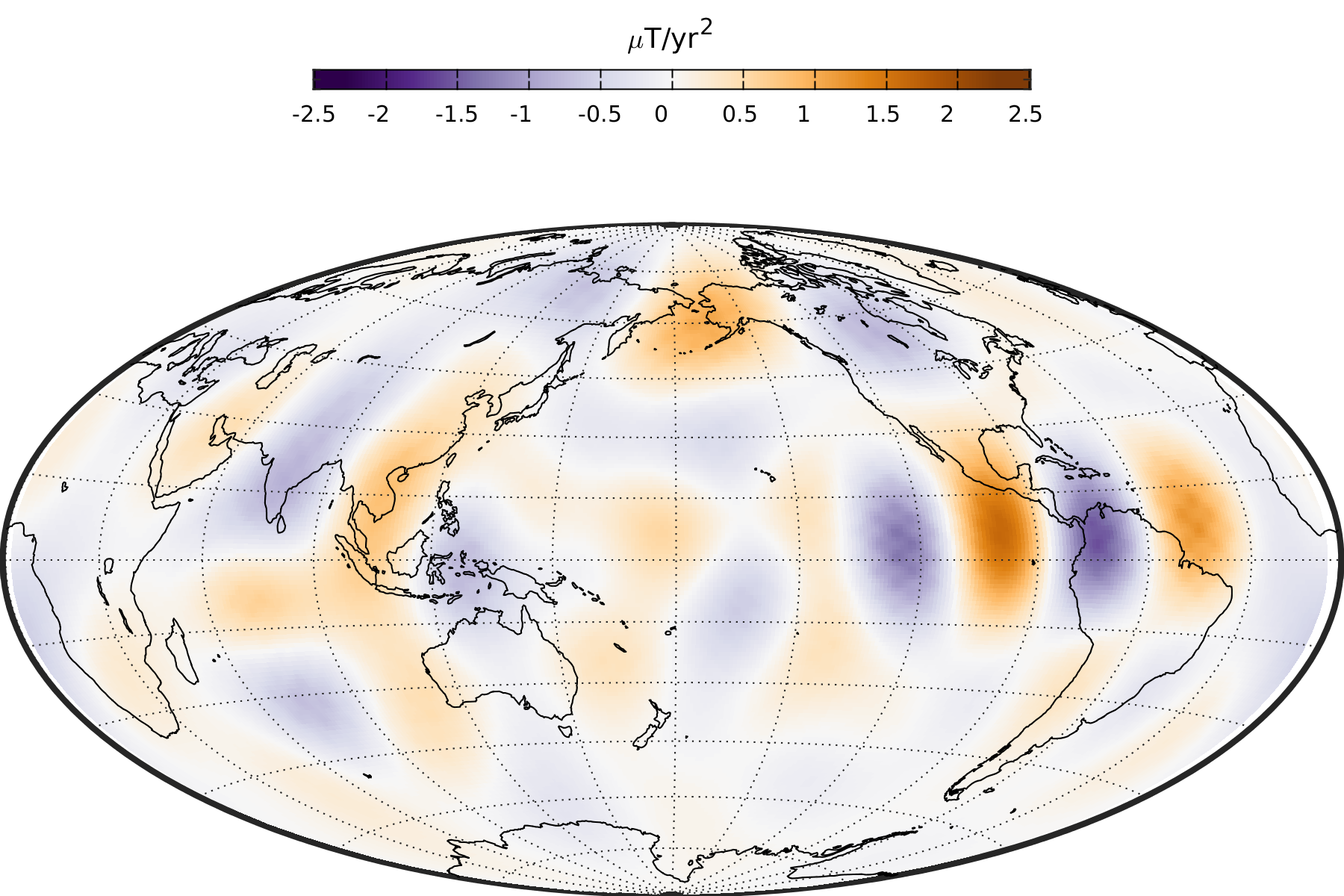}}} 
\caption{Maps of radial SA at the CMB centred on 2017.0 derived from differences of SOLA SV estimates in 2016.0 and 2018.0 using data from 2015-2017 and 2017-2019.}
\label{Fig:10}
\end{figure*}

Next, we go further and investigate the second time derivative or secular acceleration (SA) of the radial field at the CMB, which is of great interest for investigating the dynamics of the core \cite[e.g.][]{Finlay_etal_2016a, ChiDuran_etal_2020}. Again we first compare maps based on CryoSat-2 and {\it Swarm} data.  To obtain SA estimates we initially use the accumulated change between SV estimates two years apart.  In particular, in Figure \ref{Fig:10} we show the SA in 2017 based on the difference between SOLA SV estimates in 2016.0 and 2018.0. Note, that this is not an instantaneous secular acceleration but a centered difference in SV estimates 2 years apart, each based on 2 years of data. To study the SA, we computed SOLA SV estimates from {\it Swarm} data taking $\lambda=1 \times 10^{-2}\ \mathrm{nT}^{-1}$, and used the resulting associated averaging kernels as the target kernels for the SOLA estimates from both {\it Swarm} and CryoSat-2 data.

Figure \ref{Fig:10} presents global grids of SOLA CMB radial field SA estimates centered on 2017.0, again with a $1^{\circ}$ spacing, derived from CryoSat-2 (left plot) and {\it Swarm} (right plot) data subsets. Here the map is centered on the Pacific region where there has been interesting SA activity during the past 6 years \citep{Finlay_etal_2016a}. As for the SV maps, we find error estimates, computed assuming the contributing SV estimates have independent errors, are fairly homogeneous, with values ranging between $0.11-0.27$\replaced{$\mu\mathrm{T}/\mathrm{yr}^2$}{$\mathrm{nT}^2/\mathrm{yr}^2$} for the estimates derived using CryoSat-2 data and $0.06-0.08$\replaced{$\mu\mathrm{T}/\mathrm{yr}^2$}{$\mathrm{nT}^2/\mathrm{yr}^2$} for the estimates derived using {\it Swarm} data. In both cases kernel widths are close to $\approx42^{\circ}$, except in the auroral region. Comparing the CyroSat-2 and \textit{Swarm}-based SA maps, similar features can be observed. In particular, this is the case for the features seen under Asia and Indonesia. A distinctive feature reproduced in both maps is the sequence of intense patches of SA at low latitude\added{s} in a localized region below central America\added{, having amplitudes of approximately $1.9\pm 0.3$\replaced{$\mu\mathrm{T}/\mathrm{yr}^2$}{$\mathrm{nT}/\mathrm{yr}^2$} and $1.7\pm 0.1$\replaced{$\mu\mathrm{T}/\mathrm{yr}^2$}{$\mathrm{nT}/\mathrm{yr}^2$} for the CryoSat-2 and {\it Swarm} maps, respectively.} The location and amplitude of these features are similar in the two maps, confirming that CryoSat-2 data can be used to track such SA structures. Because the SOLA estimates are local averages, distant high latitude measurements, where ionospheric electrical currents may be prominent even during dark quiet times, will have little influence on such SV and SA estimates at low latitudes. \added{Finally, we note a strong patch under the Bering Sea of amplitude approximately $1.4\pm0.3$\replaced{$\mu\mathrm{T}/\mathrm{yr}^2$}{$\mathrm{nT}/\mathrm{yr}^2$} and $1.0\pm 0.1$\replaced{$\mu\mathrm{T}/\mathrm{yr}^2$}{$\mathrm{nT}/\mathrm{yr}^2$} for the CryoSat-2 and {\it Swarm} maps, respectively.}

%******************************************
%******************************************

\begin{figure*}[!htbp]
\setcounter{subfigure}{0}
\centerline{
\subfloat[SOLA SA, \added{2yr time window,} 2 month steps.]{\includegraphics[angle=0, width=0.5\textwidth]{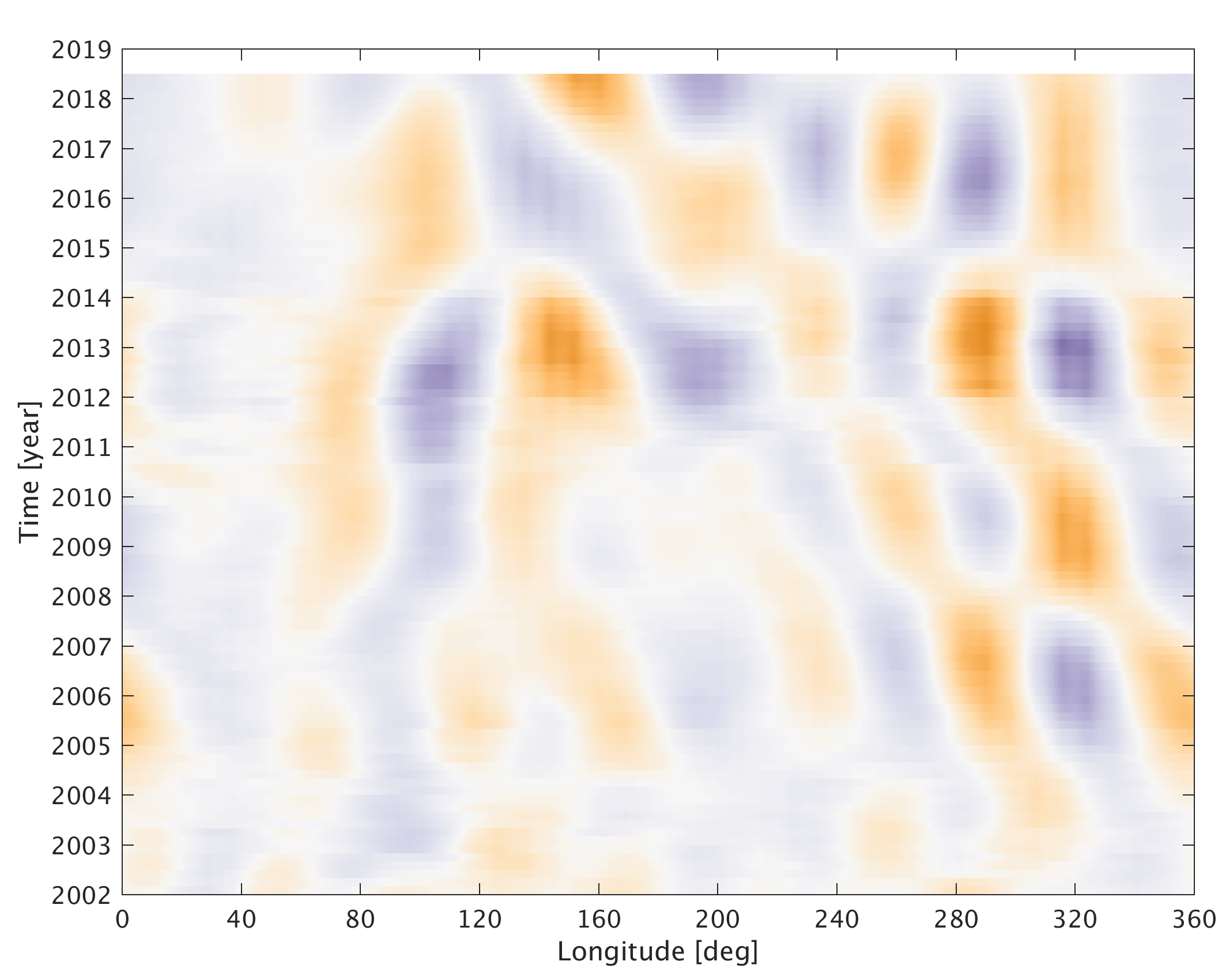}}
\hspace{0.1cm} 
\subfloat[CHAOS-7 SA, SH degrees up to 8.]{\includegraphics[angle=0, width=0.5\textwidth]{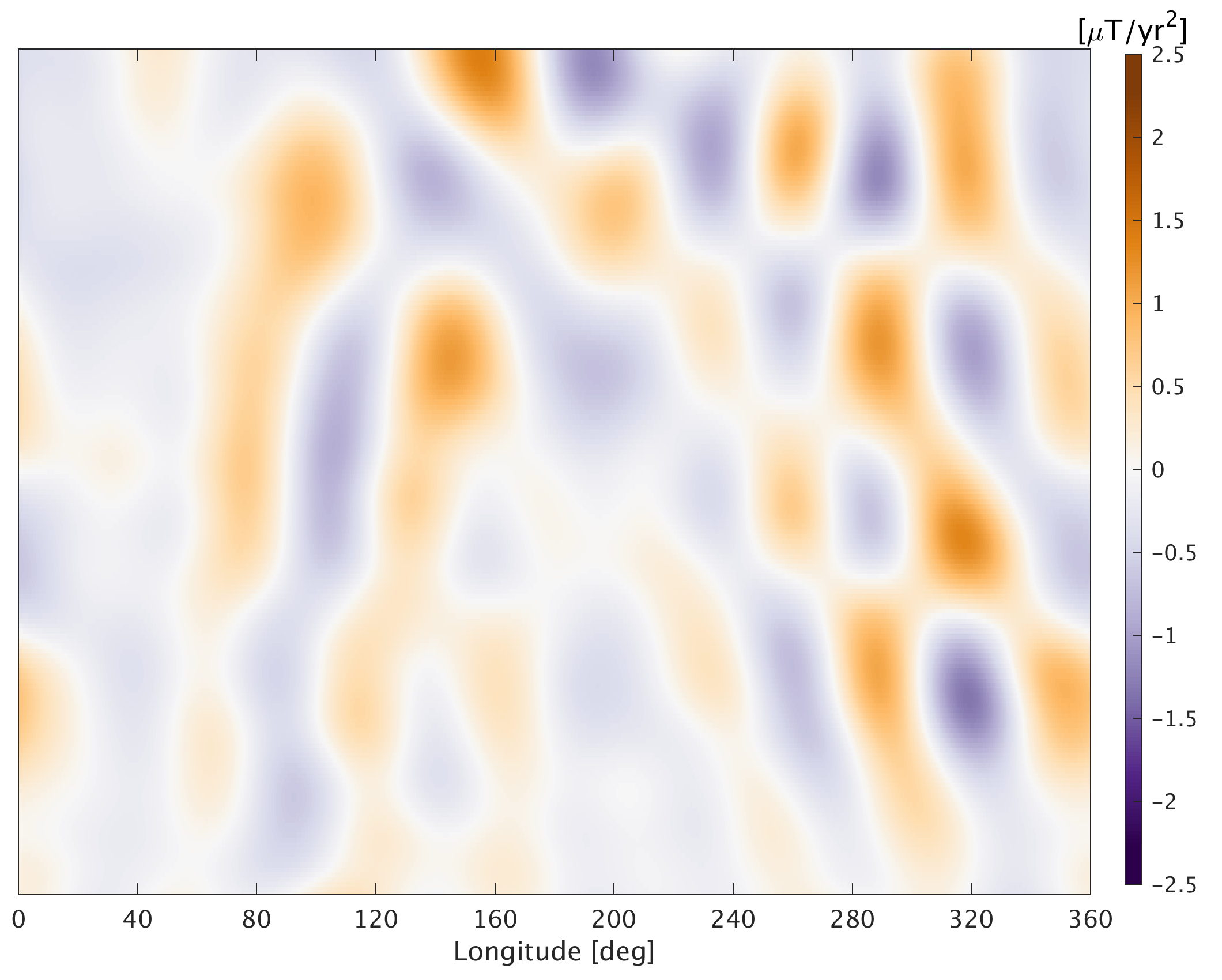}}}
\vspace{0.1cm} 
\centerline{
\subfloat[SOLA SA,  \added{2yr time window,} 2 month steps.]{\includegraphics[angle=0, width=0.5\textwidth]{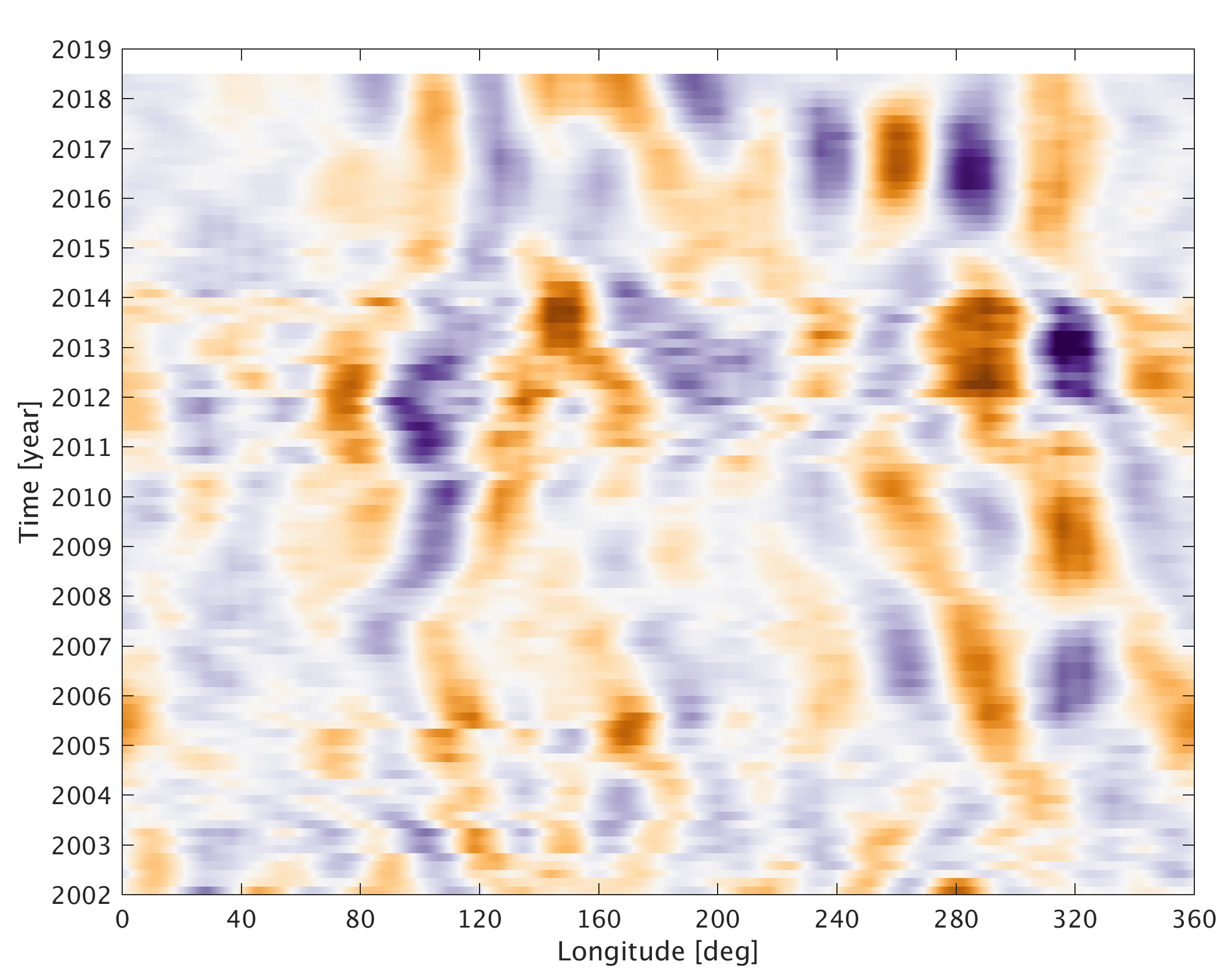}} 
\hspace{0.1cm}
\subfloat[CHAOS-7 SA, SH degrees up to 10.]{\includegraphics[angle=0, width=0.5\textwidth]{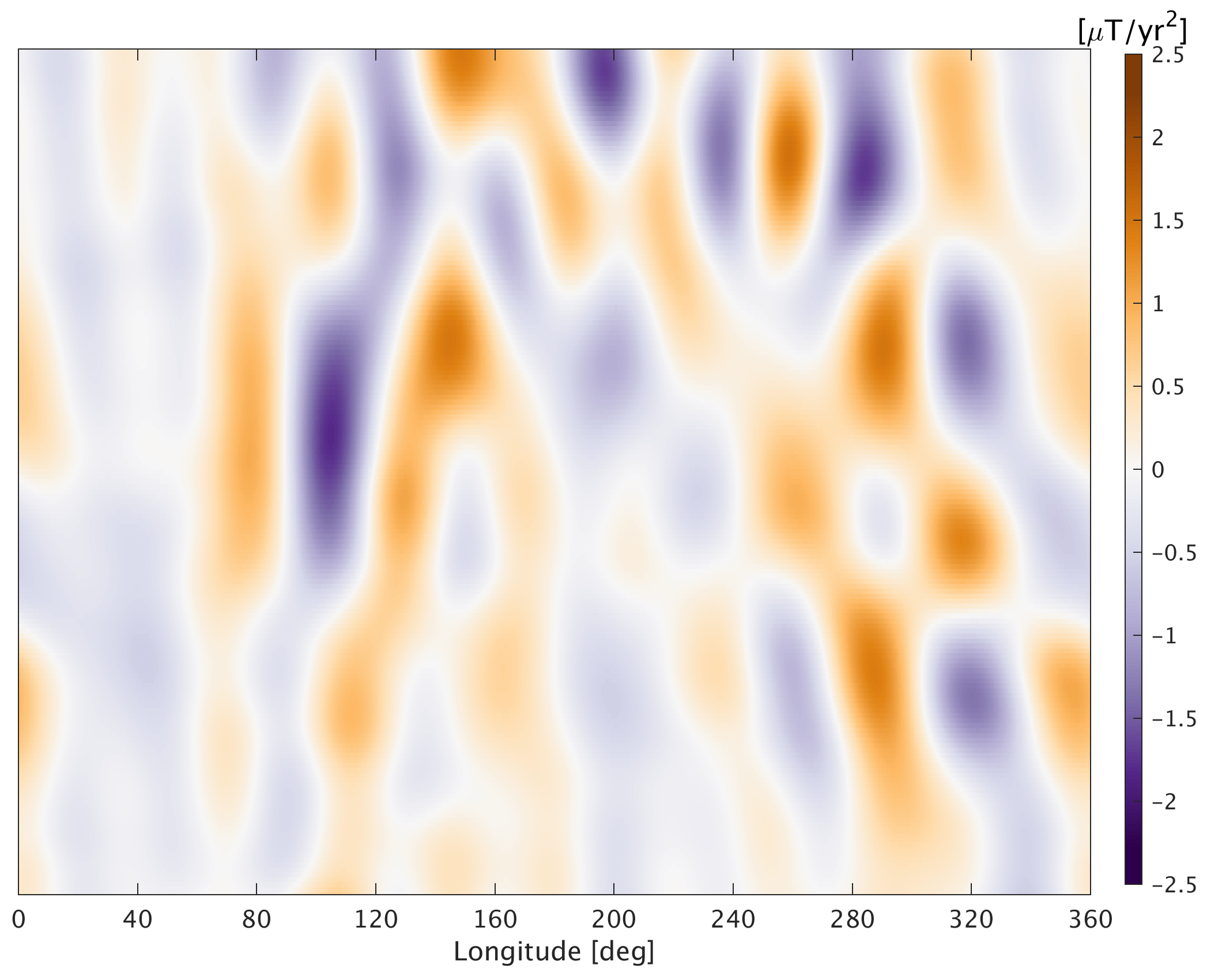}}}
\caption{Time-longitude plots of the SA along the geographical equator at the CMB. Left column: SOLA radial field SA estimates derived from differences between consecutive SOLA SV estimates based on 2yr data windows moving in 2 month steps and using $\lambda=3 \times 10^{-2}\ \mathrm{nT}^{-1}$ (top left), and $\lambda=1 \times 10^{-2}\ \mathrm{nT}^{-1}$ (bottom left). Right column: SA predictions from the CHAOS-7.2 model for SH degrees up to 8 (top right) and 10 (bottom right); note that temporal regularization causes significant time-averaging of the high degree SA in the CHAOS model.}
\label{Fig:11}
\end{figure*}

These initial investigations of the CMB radial field SV and SA using the SOLA technique indicate, as seen earlier in the GVO time series, that low latitude regions experience significant sub-decadal core field variations.  We are therefore motivated to study the field time-dependence in the equatorial region in more detail. We do this in Figure \ref{Fig:11} by constructing time-longitude (TL) plots of SOLA CMB radial field SA estimates along the geographic equator  from 2002 to 2019, centered on the Pacifc. We again compute our SA estimates based on differences of SOLA SV estimates two years apart, each derived from 2 year data windows, and sliding the windows in 2 month steps. Here we use radial field data from the {\O}rsted, CHAMP, CryoSat-2 and {\it Swarm} satellites. 

The top left plot in Figure~\ref{Fig:11} presents estimates based on averaging kernels obtained from \textit{Swarm} data using $\lambda=3 \times 10^{-2}\ \mathrm{nT}^{-1}$.  These associated error estimates between $0.02-0.08$\replaced{$\mu\mathrm{T}/\mathrm{yr}^2$}{$\mathrm{nT}^2/\mathrm{yr}^2$} and kernel widths $\approx 50^{\circ}$. Their resolution is lower that shown in Figure \ref{Fig:10} and corresponds to approximately SH degree 8. For comparison the SA predicted by the CHAOS-7.2 model for SH degrees $n \in[1,8]$ are shown on the top right plot. Note here that the CHAOS-7.2 model makes use of uncalibrated vector magnetic data from CryoSat-2 between 2010 and 2014, and co-estimates magnetometer calibration parameters \citep{Finlay_etal_2020}.  The SOLA and CHAOS TL plots in the top row of Figure \ref{Fig:11} show largely the same SA features, illustrating the convergence of the two techniques at long wavelengths of the SA and when constructing SOLA SA estimates from SV differences between consecutive 2 year time windows. For instance, the evolution of the SA features observed in Figure \ref{Fig:10} under the central Americas, can be identified ranging from longitudes $240^{\circ}$ to $320^{\circ}$ centered on 2017. In addition, we find strong SA patches in the CryoSat-2 data around 2013 at longitudes $70^{\circ}$ to $160^{\circ}$ and $280^{\circ}$ to $320^{\circ}$. Notice, that there seems to be a sign changing sequence occurring at longitudes $240^{\circ}$ to $320^{\circ}$ going from 2005 to 2019. 

Next, we increase the spatial resolution by instead deriving SV estimates using $\lambda=1 \times 10^{-2}\ \mathrm{nT}^{-1}$, which leads to slightly larger error estimates in the range $0.1-0.5$\replaced{$\mu\mathrm{T}/\mathrm{yr}^2$}{$\mathrm{nT}^2/\mathrm{yr}^2$} and kernel widths $\approx 42^{\circ}$, i.e. similar to Figure \ref{Fig:10}. This is shown in the bottom left plot while the CHAOS-7.2 model predictions for SH degrees $n \in[1,10]$, matching approximately the kernel width, are shown on the bottom right plot. Although the SOLA TL-plot looks somewhat noisier than the CHAOS plot, similar coherent evolving structures having higher amplitudes can clearly be identified. The noiser appearance in the interval 2010-2014 likely indicates the limitations of the CryoSat-2 data, but they clearly provide useful information during this period.
%******************************************
%******************************************
\newpage

\begin{figure*}[htbp]
\setcounter{subfigure}{0}
\centerline{
\subfloat[SOLA SA\replaced{, 1yr time window,}{using} 1 month steps.]{\includegraphics[angle=0, width=0.5\textwidth]{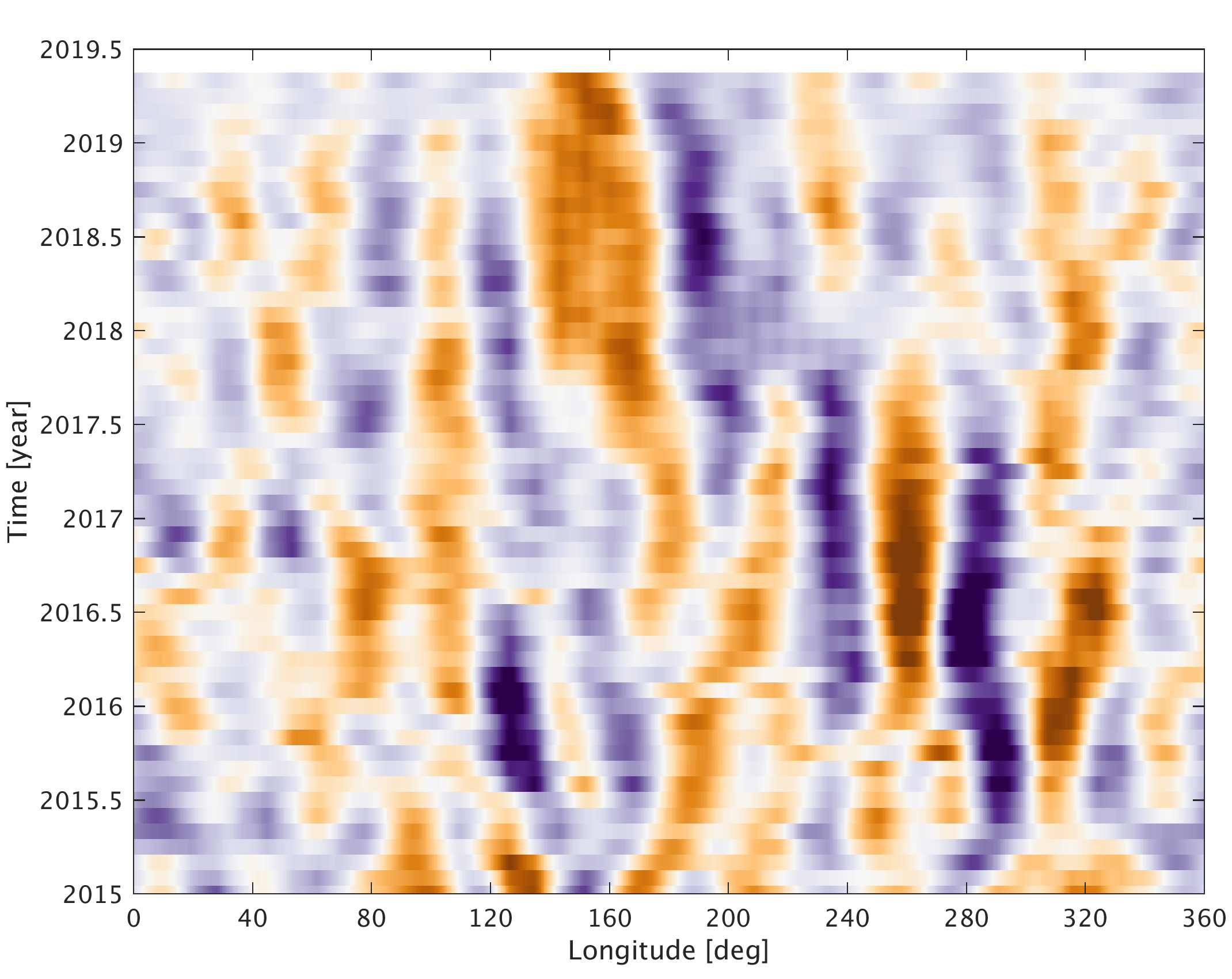}}\hspace{0.1cm} 
\subfloat[CHAOS-7 SA, SH degrees up to 10.]{\includegraphics[angle=0, width=0.5\textwidth]{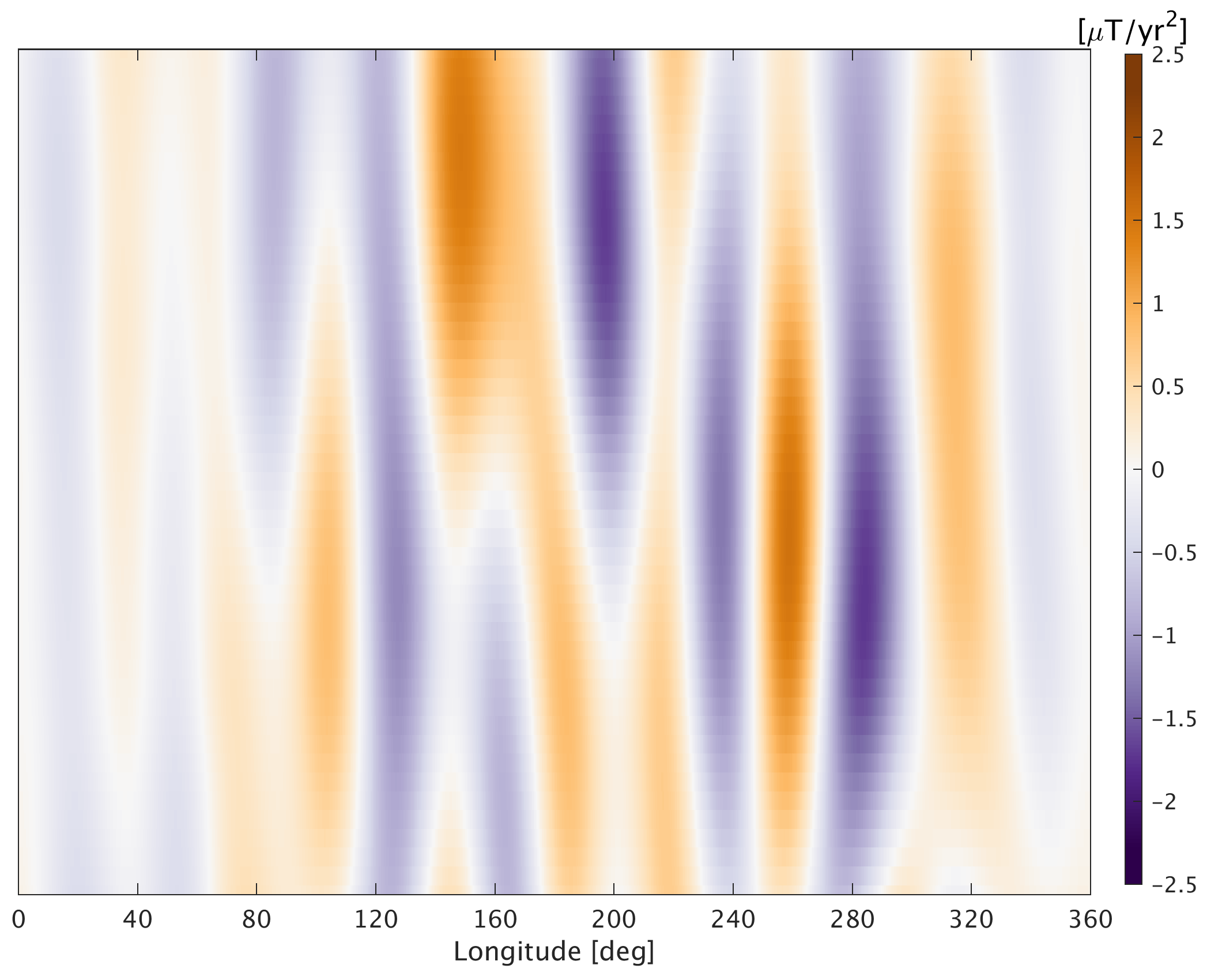}}}
\caption{Time-longitude plots along the geographical equator at the CMB, showing: 1yr differences of SOLA SV estimates derived from 1yr data windows moving in 1 month steps (left), and SA predictions from the CHAOS-7.2 model for SH degrees $n \in [1,10]$ (right).}
\label{Fig:12}
\end{figure*}

With data from the {\it Swarm} mission, it is possible to go further and also increase the temporal resolution of the SA by taking 1 year differences of SOLA SV estimates derived from 1 year data windows sliding in 1 month steps. The result of applying this procedure to obtain SA estimates on the geographic equator between 2015.0 and 2019.5 is shown in the left plot of Figure \ref{Fig:12}. These SOLA estimates have associated errors of $0.3-0.6$\replaced{$\mu\mathrm{T}/\mathrm{yr}^2$}{$\mathrm{nT}^2/\mathrm{yr}^2$} and kernel widths $\approx 42^{\circ}$. The right plot shows similar CHAOS-7.2 model predictions for SH degrees $n \in[1,10]$. Both TL-plots shows the similar large scale features, for instance, the features under central America from longitudes $240^{\circ}$ to $320^{\circ}$, which are elongated compared with Figure~\ref{Fig:11} due to the change in scale of the y-axis (time). However the SOLA results show significantly more time dependence, revealing features that were smoothed out by the temporal regularization of CHAOS-7.2.  Changes of sign in the SA within about 1 year can be observed.  
%******************************************
%******************************************

Particularly interesting is the appearance in the Pacific region around 2017, at longitudes $150^{\circ}$ to $220^{\circ}$, of side-by-side positive and negative intense SA features, that have subsequently drifted westwards. This SA change coincides with the peak in the radial SV field observed in the Pacific region during {\it Swarm} time seen in the GVO map Figure \ref{Fig:4}.  We note the presence of features in Figure~\ref{Fig:12} that appear to drift rapidly both eastwards and westwards, for example from 160$^\circ$ East in 2015 to 220$^\circ$ East in 2017. \added{Such rapidly drifting behavior of low latitude SA patches is difficult to explain in terms of simple core flow advection processes.  They may instead be a signature of wave propagation close to the core surface.  A range of possible candidates for fast waves in the core have recently been described, some requiring only a strong magnetic field and rotation  \mbox{\citep{Aubert_Finlay_2019, Gerick_etal:2020}} while others rely on the presence of a possible stratified layer at the top of the core \mbox{\citep{Buffett_Matsui_2019}}.} Though tempting, it may be dangerous to interpret such features that are at the limit of the present spatial resolution and temporal resolution \citep{Gillet_2019}. It will be important to assess whether such features remain coherent in the future, as the resolution of the SA increases.

\section{Conclusions}
\label{sec:Conclusions}
% What have done
In this article we have studied global patterns and sub-decadal changes in geomagnetic secular variation during the past 20 years.  We have shown there is now continuous coverage of magnetic field measurements from low Earth orbiting satellite missions during this period, provided one takes advantage of calibrated platform magnetometer data from the CryoSat-2 satellite. Using vector magnetic field measurements from {\O}rsted, CHAMP, Cryosat-2 and \textit{Swarm} we have constructed Geomagnetic Virtual Observatory (GVO) time series that track sub-decadal changes in the core field secular variation at satellite altitude, and we have used the Subtractive Optimally Localized Averages (SOLA) technique to study  the secular variation of the radial field and its time changes down at the core-mantle boundary. These are both local methods whereby field measurements in the vicinity of the location of interest are combined so as to estimate the field at that point; measurements far away from the site of interest, have little or no influence on the field estimates. 

% Extra info/insight
Using the GVO method, we derived composite time series of geomagnetic secular variation, spanning nearly 20 years, on a global grid of 300 GVOs. GVO time series derived from CHAOS-6-x9 and IGRF-13 calibrated CryoSat-2 data, show similar sub-decadal SV features, and comparable level of scatter.  We found a scatter level for radial field SV of $3.5$~nT/yr for the CryoSat-2 GVOs compared with $2.5$~nT/yr for CHAMP and $1.0$~nT/yr for {\it Swarm}. Comparing GVOs with overlapping epochs from 2014 to 2018, derived from CHAOS-6x9 calibrated CryoSat-2 data and {\it Swarm} data, we find similar sub-decadal SV changes, thus confirming the possibility of using CryoSat-2 for core field studies. In our 20~yr long composite record, we observe fluctuations in the radial SV field of up to 20~nT/yr in amplitude occurring at low latitudes over time periods of 5-10 years. For instance, we see a rapid change of slope over Indonesia around 2014, over the South America and South Atlantic region around 2007, 2011 and 2014, and in the central Pacific around 2017. Some of these events have previously been discussed in ground observatory records \citep{Brown_etal_2013,Torta_etal_2015}. They have the distinct "$\Lambda$" and "$V$"-shapes that are often associated with geomagnetic jerks, but as indicated in earlier investigations \citep{Olsen_Mandea_2007,Olsen_Mandea_2008, Chulliat_Maus_2014} we find these events are localized rather than global features.

Using the SOLA technique, we mapped the radial field SV directly at the core-mantle boundary using satellite data, computing spatially localized averages of the SV, time-averaged over specified windows.  Taking differences between consecutive SV estimates we obtained estimates for the secular acceleration (SA) at the CMB.  Using only CryoSat-2 measurements we are able to successfully map the SA at the CMB, down to spatial averaging widths of $\approx42^{\circ}$, corresponding approximately to SH degree 10 or length scales of 2500\,km \added{at the CMB}. Comparing SV and also SA field maps at the CMB, derived from CryoSat-2 and {\it Swarm} measurements, the same features can be identified having similar amplitudes and latitude/longitude extend. 

In time-longitude plots of SOLA-based radial field SA estimates along the geographic equator at the CMB we find strong SA features, with amplitudes of $\pm 2.5$\replaced{$\mu\mathrm{T}/\mathrm{yr}^2$}{$\mathrm{nT}^2/\mathrm{yr}^2$}, under Indonesia at longitudes $70^{\circ}$ to $160^{\circ}$ from 2011 to 2014 during CryoSat-2 time, under central America at longitudes $240^{\circ}$ to $320^{\circ}$ appearing from 2015 to 2019, and in sequences with alternating SA signs under South America and the South Atlantic region at longitudes $240^{\circ}$ to $360^{\circ}$ during 2004-2019. The imaged SA features around 2013 at longitudes $70^{\circ}$ to $160^{\circ}$ and $280^{\circ}$  $360^{\circ}$ demonstrates the usefulness of the CryoSat-2 measurements.  Our results lend support to a sign changing sequence of SA, for observable length scales down to 2500 km in this study, at longitudes $240^{\circ}$ to $360^{\circ}$ from 2005 to 2019, that has been noticed in previous studies \citep{Chulliat_etal_2015,Alken_etal_2020b}.
We have shown it is possible to increase the temporal resolution of SA estimates during {\it Swarm} era, compared to that seen for example in the CHAOS-7 model, by computing SA estimates from the differences of consecutive SOLA SV estimates derived using 1-yearly time windows. We find the similar coherent structures as seen in TL plots constructed using 2-yearly time window, but also see changes of sign in the SA within 1 year. In the central Pacific region at longitudes $150^{\circ}$ to $220^{\circ}$ we find strong positive and negative SA features appearing side-by-side in late 2017 drifting westwards until 2020. \added{The results presented in Figure \ref{Fig:11} and \ref{Fig:12}, demonstrate that estimates of core field SA different from those found in the CHAOS model can be obtained despite similar data selection and external field modelling schemes.  This is due to the important role played by the model parameterization and regularization in the SA recovered in the CHAOS model, especially for the small length-scales and fast time changes which are at the limit of what can be reliably resolved from the data.}

% Geophysical implications/interpretations
The rapid fluctuations of the core magnetic field described in this study are likely caused by time variations in the motions of the liquid metal outer core. In particular, changes in secular acceleration patterns at low latitudes provide constraints on the equatorial dynamics of the outer core \citep{Aubert_Finlay_2019,Kloss_Finlay_2019}. This is a topic of active research with various phenomenon recently proposed including equatorially-trapped MAC waves in a stratified layer close to the core surface \citep{Buffett_Matsui_2019,ChiDuran_etal_2020,Gerick_etal:2020} or the equatorial focusing of hydrodynamic waves driven by turbulent convection deep within the core   \citep{Aubert_Finlay_2019}. 

% Future possibilities
It is undeniable that much core dynamics occurs on time-scales either much longer, or much shorter,  than can be can be resolved using the current available satellite and ground observations \citep{Gillet_2019}. However, with 20 years of low-Earth orbit satellite measurements of the vector magnetic field, and with tools similar to those presented here, we are now able to probe and characterize core field changes with increasing spatial and temporal resolution. We have shown here that platform magnetometer data can help in this activity, provided they are appropriately calibrated.

\begin{backmatter}

\section*{List of abbreviations}
AT - Along-track,
CHAMP - CHAllenging Minisatellite Payload,
CHAOS - CHAMP, {\O}rsted, and \textit{Swarm} field model,
CI - Comprehensive Inversion Model,
CMB - Core-Mantle Boundary,
ECEF - Earth-Centered-Earth-Fixed reference frame
$E_m$ - Merging Electric Field at the Magnetopause,
EW - East-West,
FGM - Fluxgate Magnetometer,
GCV - Generalized cross-validation,
GVO - Geomagnetic Virtual Observatory,
IGRF - International Geomagnetic Reference Field,
IMF - Interplanetary Magnetic Field,
Kp - K planetary index,
LCS - Lithospheric model from CHAMP and Swarm,
LEO - Low Earth orbiting,
nT - nano-Tesla,
QD - Quasi-Dipole,
RC - Ring-Current index,
rms - Root-Mean Square,
RMM - Revised Monthly Mean,
SA - Secular Acceleration,
SH - Spherical Harmonic,
SOLA - Subtractive Optimally Localized Averages,
SV - Secular Variation,
TL - Time longitude,
yr - year.

\section*{Declarations}

\section*{Availability of datasets and material}
GVO datasets derived from the {\it Swarm}, CHAMP and CryoSat-2 missions, along with software and additional documentation are available from the GVO project webpage, \url{https://www.space.dtu.dk/english/research/projects/project-descriptions/geomagnetic-virtual-observatories}\\
The CryoSat-2 datasets are available at swarm-diss.eo.esa.int in folder \#CryoSat-2.

\section*{Competing interests}
The authors declare that they have no competing interests.

\section*{Funding}
MDH and CCF were funded by the European Research Council (ERC) under the European Union’s Horizon 2020 research and innovation programme (grant agreement No. 772561). The study has been partly supported by the Swarm+ 4D Deep Earth: Core project, ESA contract no. 4000127193/19/NL/IA and as part of \textit{Swarm} DISC activities, funded by ESA contract no. 4000109587.

\section*{Author's contributions}
MDH developed the GVO data processing and modelling scheme and drafted the manuscript. MDH and CCF developed the SOLA modelling scheme.  NiO calibrated and prepared the CryoSat-2 data and developed the original version of the GVO modelling scheme.  All authors contributed to the design of the study. All authors read and approved the final manuscript.

\section*{Acknowledgements}
We thank the GFZ German Research Centre for Geoscience for providing access to the CHAMP MAG-L3 data and the European Space Agency (ESA) for providing access to the CryoSat-2 and the {\it Swarm} L1b data. High resolution 1-min OMNI data was provided by the Space Physics Data Facility (SPDF), NASA Goddard Space Flight Center. We like to thank two anonymous reviewers for comments that helped improve the manuscript.

%\section*{Authors details}
%Address

% if your bibliography is in bibtex format, use those commands:
%\bibliographystyle{aps-nameyear}   
\bibliographystyle{agu08}   
\bibliography{EPS_2019}      % Bibliography file (usually '*.bib' )

\clearpage

\end{backmatter}
\end{document}